\documentclass[twocolumn]{aastex61}
\usepackage{amsmath}
\usepackage{amssymb}
\usepackage{latexsym}
\usepackage{listings}
\usepackage{theorem}
\usepackage{verbatim}
\usepackage{multirow}
\usepackage{natbib}
\usepackage{hhline}
\DeclareGraphicsExtensions{.pdf}
\shorttitle{False-Positives in WISE Circumstellar Disk Surveys}
\shortauthors{Silverberg et al.}

\begin{document}


\title{Follow-up Imaging of Disk Candidates from the Disk Detective Citizen Science Project: New Discoveries and False-Positives in WISE Circumstellar Disk Surveys}
\author{Steven M. Silverberg}
\affiliation{Homer L. Dodge Department of Physics and Astronomy, University of Oklahoma, 440 W. Brooks Street, Norman, OK 73019, USA}
\affiliation{NASA Goddard Space Flight Center, Exoplanets and Stellar Astrophysics Laboratory, Code 667, Greenbelt, MD 20771, USA}

\author{Marc J. Kuchner}
\affiliation{NASA Goddard Space Flight Center, Exoplanets and Stellar Astrophysics Laboratory, Code 667, Greenbelt, MD 20771, USA}

\author{John P. Wisniewski}
\affiliation{Homer L. Dodge Department of Physics and Astronomy, University of Oklahoma, 440 W. Brooks Street, Norman, OK 73019, USA}

\author{Alissa S. Bans}
\affiliation{Department of Physics, Emory University, 201 Dowman Drive, Atlanta, GA 30322, USA}

\author{John H. Debes}
\affiliation{Space Telescope Science Institute, 3700 San Martin Dr., Baltimore, MD 21218, USA}

\author{Scott J. Kenyon}
\affiliation{Smithsonian Astrophysical Observatory, 60 Garden Street, Cambridge, MA 02138, USA}

\author{Christoph Baranec}
\affiliation{Institute for Astronomy, University of Hawai`i at M\={a}noa, Hilo, HI 96720-2700, USA}

\author{Reed Riddle}
\affiliation{Division of Physics, Mathematics, and Astronomy, California Institute of Technology, Pasadena, CA 91125, USA}

\author{Nicholas Law}
\affiliation{Department of Physics and Astronomy, University of North Carolina at Chapel Hill, Chapel Hill, NC 27599-3255, USA}

\author{Johanna K. Teske}
\affiliation{Carnegie DTM, 5241 Broad Branch Road, NW, Washington, DC 20015, USA}

\author{Emily Burns-Kaurin}
\affiliation{Disk Detective Advanced User Team}

\author{Milton K.D. Bosch}
\affiliation{Disk Detective Advanced User Team}

\author{Tadeas Cernohous}
\affiliation{Disk Detective Advanced User Team}

\author{Katharina Doll}
\affiliation{Disk Detective Advanced User Team}

\author{Hugo A. Durantini Luca}
\affiliation{Disk Detective Advanced User Team}

\author{Michiharu Hyogo}
\affiliation{Disk Detective Advanced User Team}

\author{Joshua Hamilton}
\affiliation{Disk Detective Advanced User Team}

\author{Johanna J. S. Finnemann}
\affiliation{Disk Detective Advanced User Team}

\author{Lily Lau}
\affiliation{Disk Detective Advanced User Team}

\collaboration{Disk Detective Collaboration}

\correspondingauthor{Steven M. Silverberg}
\email{silverberg@ou.edu}

\begin{abstract}
The Disk Detective citizen science project aims to find new stars with excess 22-$\mu$m emission from circumstellar dust in the AllWISE data release from the Wide-field Infrared Survey Explorer (WISE). We evaluated 261 Disk Detective objects of interest with imaging with the Robo-AO adaptive optics instrument on the 1.5m telescope at Palomar Observatory and with RetroCam on the 2.5m du Pont telescope at Las Campanas Observatory to search for background objects at $0.15''-12''$ separations from each target. Our analysis of these data lead us to reject $7\%$ of targets. Combining this result with statistics from our online image classification efforts implies that at most 7.9\% $\pm$ 0.2\% of AllWISE-selected infrared excesses are good disk candidates. Applying our false positive rates to other surveys, we find that the infrared excess searches of \citet{2012MNRAS.427..343M}, \citet{2017MNRAS.471..770M}, and \citet{2016MNRAS.458.3479M} all have false positive rates $>70\%$. Moreover, we find that all thirteen disk candidates in \citet{2014ApJ...794..146T} with W4 signal-to-noise $>3$ are false positives. We present 244 disk candidates that have survived vetting by follow-up imaging. Of these, 213 are newly-identified disk systems. Twelve of these are candidate members of comoving pairs based on \textit{Gaia} astrometry, supporting the hypothesis that warm dust is associated with binary systems. We also note the discovery of 22 $\mu$m excess around two known members of the Scorpius-Centaurus association, and identify known disk host WISEA J164540.79-310226.6 as a likely Sco-Cen member. Thirty-one of these disk candidates are closer than $\sim 125$ pc (including 27 debris disks), making them good targets for direct imaging exoplanet searches. 




\end{abstract}

\section{Introduction}
\label{sec:introduction}


With higher sensitivity than any previous full-sky infrared survey instrument, the \textit{Wide-field Infrared Survey Explorer} \citep[WISE;][]{2010AJ....140.1868W} detected over $747$ million sources in its all-sky survey. Many teams have searched for new circumstellar disks in the WISE data based on infrared excess at W4 (22 $\mu$m emission) compared to W1 (3.4 $\mu$m emission), discovering thousands of candidate debris disks \citep[see e.g.][]{2013MNRAS.433.2334K, 2013ApJS..208...29W, 2014ApJS..212...10P, 2016ApJS..225...15C, 2017AJ....153...54P} and YSO disks \citep{2011ApJ...733L...2L, 2011ApJS..196....4R, 2014AJ....147..133L, 2014ApJ...791..131K, 2015AJ....150..100K}. In particular, because of its sensitivity and full-sky scope, the WISE mission is uniquely suited to the search for M dwarf debris disks, which are of particular interest due to the relative lack of detected disks around these stars in comparison to higher-mass stars \citep[e.g.][]{Plavchan2005,Plavchan2009,Lestrade2009,Avenhaus2012,2014ApJ...794..146T,MoreyLestrade2014,BinksJeffries2017}.

However, confusion and contamination limit every search for disks with WISE. The point spread function (PSF) at W4 has a full-width at half-maximum of $12''$. This wide PSF can allow emission from multiple point sources (e.g., background stars) or image artifacts to contribute to the W4 photometry, producing a false-positive [W1]-[W4] excess. Additionally, the color loci of debris disks {and YSOs} overlaps the color loci of several other astronomical phenomena, including background galaxies and stars embedded in nebulosity \citep{2012ApJ...744..130K}. To eliminate these false-positives, objects must be examined in visible and near-infrared images along with the WISE images. Most published searches have utilized visual inspection of the WISE images \citep[see e.g.][]{2011ApJ...729....4D, 2012MNRAS.426...91K, 2013ApJS..208...29W, 2014MNRAS.437..391C, 2014ApJS..212...10P} to address these contamination and confusion problems.

\citet{2012MNRAS.426...91K} produced a study of these confusion and contamination issues, focusing on infrared excesses around stars observed by \textit{Kepler} \citep{2010Sci...327..977B}, with the goal of expanding the number of known stars that host both planets and debris disks. They searched for infrared excesses in a cross-match of the Kepler Input Catalog (KIC) to 2MASS and WISE, finding 7965 disk candidates. However, they argued that all but 271 (3.4\%) of these objects were coincident with Galactic dust emission, as identified by the IRAS 100 $\mu$m background, and therefore false positives.

The Disk Detective citizen science/crowdsourcing project \citep[][hereafter Paper 1]{Kuchner2016} uses citizen science to examine infrared excess candidates from WISE, beginning with a website, http://www.diskdetective.org, where volunteers examine images from WISE, the Two-Micron All-Sky Survey (2MASS), the Digitized Sky Survey (DSS), and the Sloan Digital Sky Survey (SDSS) to check for false positives. Since launch in January 2014, over 30,000 users have made over 2.6 million classifications via this Zooniverse website. Shortly after project launch, a group of highly-dedicated volunteers began their own email discussion group for the project. Since then, members of this ``advanced user group'' have helped train other users and research follow-up targets in the literature, and now form a crucial extension of the Disk Detective science team (Paper 1). 

In this paper, we address two specific forms of false positive that occur in the Disk Detective input catalog and other searches for circumstellar disks with WISE:

\begin{enumerate}
    \item \textbf{Confusion in the WISE images:} the contribution to noise in an image due to superposed signals from faint unresolved sources that cluster on the scale of the observing beam.
    \begin{enumerate}
        \item We use data from the Digitized Sky Survey (DSS), the Sloan Digital Sky Survey (SDSS), and the 2-Micron All-Sky Survey (2MASS) to search for background sources via the Disk Detective website.
        \item We use high-resolution imaging on small telescopes (the 1.5-m Telescope at Mt. Palomar and the Du Pont Telescope at Las Campanas) to identify background sources that are too faint for SDSS and 2MASS or unresolved by these surveys.
    \end{enumerate}
    \item \textbf{Contamination of the WISE images:} the presence of image artifacts (e.g. diffraction spikes, latent images, or optical ghosts) within the WISE beam. We search for these contaminants by examining the WISE images via the Disk Detective website.  
\end{enumerate}

\noindent
We present the first results from our follow-up imaging campaigns with the Robo-AO adaptive optics instrument on the 1.5-m telescope at Palomar Observatory \citep{2014ApJ...790L...8B} and the RetroCam instrument on the Du Pont telescope at Las Campanas Observatory in Chile \citep[e.g.][]{2014SPIE.9147E..5LR}. These instruments  provided an angular resolution of 0\farcs 15 and $< 1''$ respectively, improving on the $5''$ effective resolution of the 2MASS Point Source Catalog\footnote{As listed here: \url{https://www.ipac.caltech.edu/2mass/releases/allsky/doc/sec2_2a.html}}. We combine these observations with the results of the website-based evaluation to estimate the fraction of WISE excesses that are true disk candidates. In Section \ref{sec:website}, we review our website classification procedure, analyze the distribution of clean sources and false positives, and consider the results of an advanced-user-driven literature review. In Section \ref{sec:target_selection}, we review our procedure for selecting these particular targets for further examination, and discuss our methodology for collecting high-resolution images of these targets. In Section \ref{sec:visual_analysis}, we describe our method for  identification of contaminated targets. In Section \ref{sec:multiple_rates}, we present the results of our follow-up observations, estimate the presence of further unresolved sources, and combine these with the results from website analysis to estimate the ultimate yield of the Disk Detective input catalog. We apply these findings to other surveys in Section \ref{sec:other_surveys}, including re-analysis of the M dwarf disk search of \citet{2014ApJ...794..146T}. In Section \ref{sec:new_candidates}, we present our list of uncontaminated disk candidates.  Finally, in section \ref{sec:Conclusion} we summarize our results and discuss future plans for these targets and high-resolution follow-up imaging.

\section{Rejection of False Positives Via the DiskDetective.org Website and Literature Review}
\label{sec:website}

The online engine of our citizen science disk search is DiskDetective.org, where users view sets of images showing the same WISE point source in several bands. In this section, we review our online classification method (discussed in more depth in Paper 1), and analyze the latest online classification results.



\subsection{DiskDetective.org: Identification of WISE Debris Disk Candidates with Citizen Science}

The selection of our input catalog is detailed in Table 1 of Paper 1. Briefly, we selected objects from the AllWISE Source Catalog with significant [W1]-[W4] excess ($\mathrm{[W1]-[W4]} > 0.25, \mathrm{[W1]-[W4]} > 5\sigma_{[W1]-[W4]}$) and high signal-to-noise (SNR $> 10$) at W4 that were not flagged as being contaminated in any way as part of the AllWISE source processing. These objects become ``subjects'' as part of our input catalog.\footnote{The full input catalog is available via the MAST archive, \url{https://mast.stsci.edu}.} For each subject, we generated a ``flipbook'' of 9-15 1-arcminute-square images of the subject from the Digitized Sky Survey (DSS), the seventh data release of the Sloan Digital Sky Survey (SDSS DR7), 2MASS, and AllWISE. Images were overlaid with a red circle of radius 10.5 arcsec, the area that must be clear of contamination for the AllWISE photometry to be trustworthy, and small crosshairs indicating the center of the W1 source. Users view each flipbook as an animation, or scroll through frame-by-frame using a scrub bar. Users then choose from six classification buttons, labeled ``Multiple objects in the Red Circle,'' ``Object Moves off the Crosshairs,'' ``Extended beyond circle in WISE Images,'' ``Empty Circle in WISE Images,'' ``Not Round in DSS2 or 2MASS images,'' and ``None of the Above/Good Candidate.'' Users can choose either ``None of the Above,'' or as many of the other descriptions as apply to a target. This classification method robustly identifies type-1a and type-2 false positives, as described above.

In this paper we focus only on subjects that have been retired from active classification, which we refer to as ``complete.'' In Paper 1, we defined this cutoff as 15 classifications. After the publication of Paper 1, we put into place a new retirement scheme, and corrected some objects for saturation effects at W1. The new retirement scheme is described in Appendix \ref{sec:retirement}, and our correction for W1 saturation is treated in Appendix \ref{sec:dropouts}. As of 2018 January 5, 62\% of subjects were complete, providing the large sample of 149,273 subjects we analyze here.



We define a ``good'' subject as one where the majority of classifiers of a subject label it ``None of the Above/Good Object.'' We refer to a subject as ``multiple'' if a majority of classifiers labels that subject as having ``Multiple objects in the Red Circle;'' these ``multiple'' subjects are the dominant false positive rejected by our volunteers. We choose this definition (based on a majority of classifiers rather than a majority of votes) because users can select more than one option with each classification; this metric shows us subjects that most users agree have ``multiple objects in the red circle,'' even if they also have other flaws. For convenience, we refer to all subjects that do not meet these definitions of ``good'' or ``multiple'' as ``other'' in this paper.


\subsection{False Positive Rates in Website Classifications}


\begin{figure*}[htb]
\begin{centering}
\plottwo{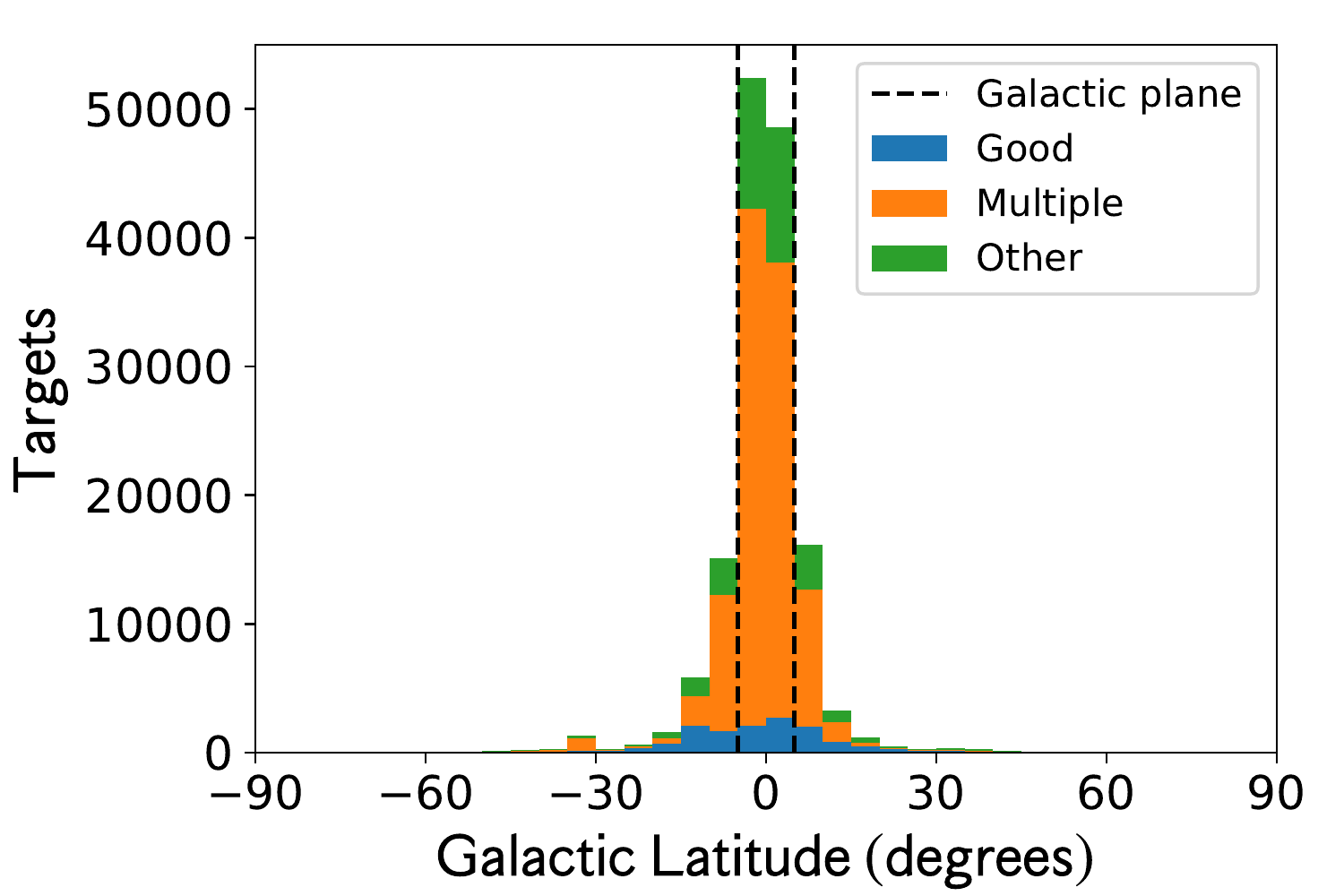}{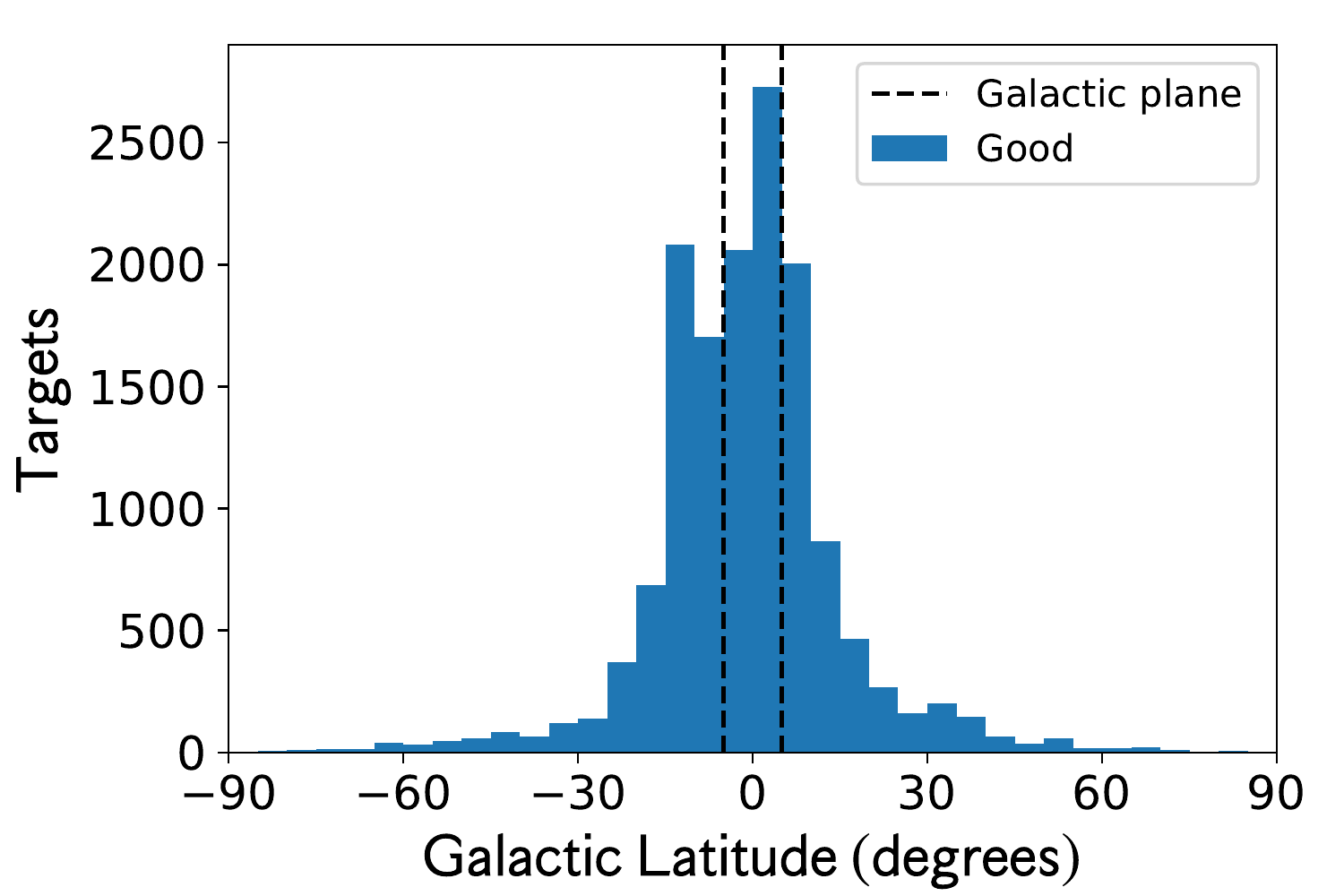}
\caption{Left: distribution of complete subjects as a function of Galactic latitude. Right: distribution of complete ``good" subjects. ``Good'' subjects were classified as ``None of the above--good candidate'' by more than $50\%$ of classifiers.``Multiple'' subjects were classified as ``Multiple objects in the red circle'' by more than $50\%$ of classifiers. All other complete subjects were labeled ``other.''}
\label{fig:classification_data_raw_numbers}
\end{centering}
\end{figure*}

\begin{figure}[htb]
\begin{centering}
\includegraphics[width=0.5\textwidth]{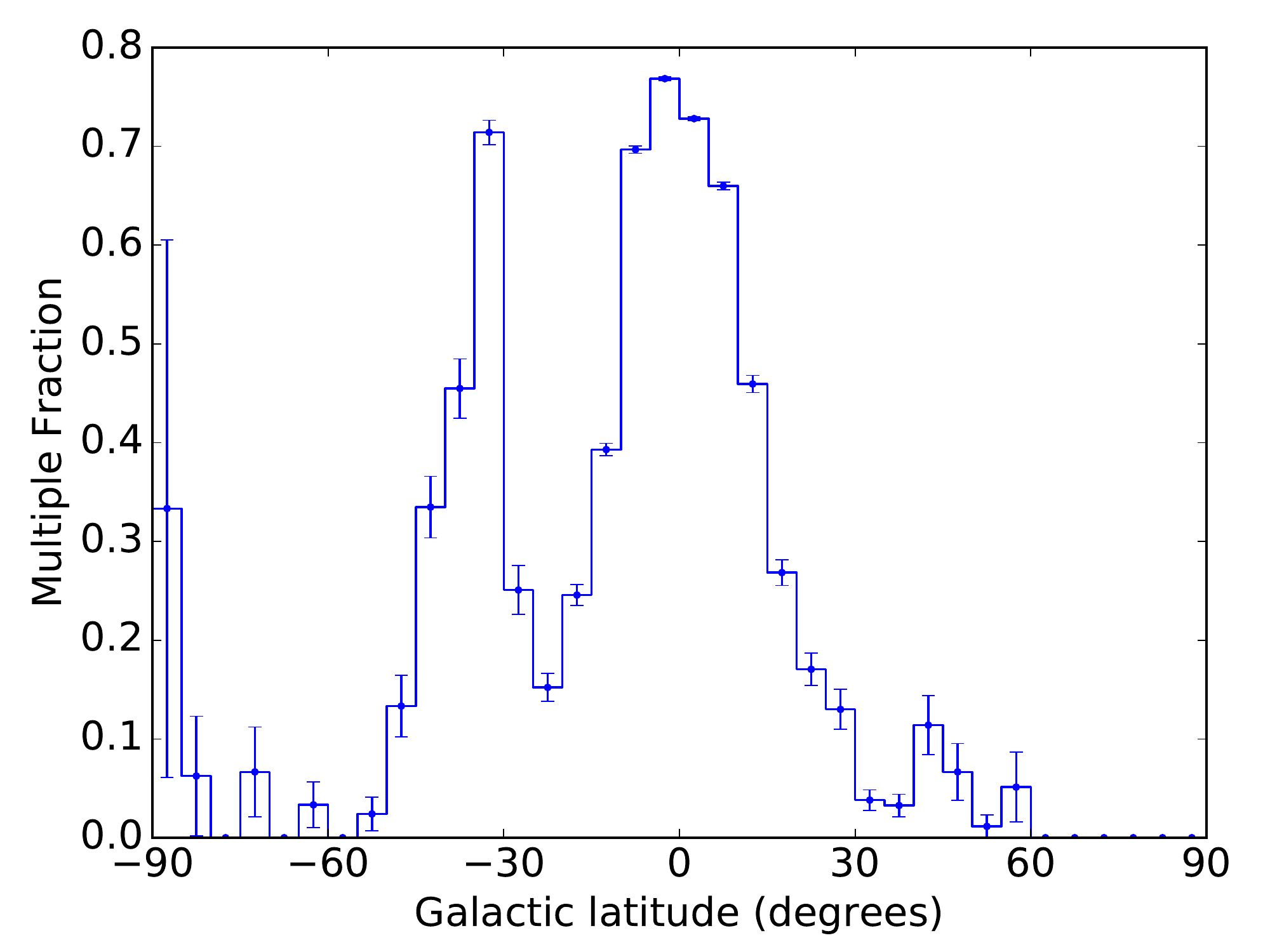}
\caption{Rate of ``multiple'' subjects as a fraction of all complete subjects as a function of Galactic latitude. ``Multiple'' subjects are most common in the Galactic plane, and in the Large Magellanic Cloud (LMC).}
\label{fig:multiple_rate}
\end{centering}
\end{figure}

Figure \ref{fig:classification_data_raw_numbers} shows the distribution of raw numbers of complete subjects broken down by category (``good,'' ``multiple,'' or ``other'') as a function of Galactic latitude, and the distribution of ``good'' objects alone as a function of Galactic latitude. Most of the complete subjects are false-positives; only $9.80\%$ ($\pm 0.08\%$) of the complete subjects are labeled ``good'' by a majority of volunteers. ``Multiples'' are the dominant form of false positive---they make up $68.87\%$ ($\pm 0.12\%$) of all complete subjects.

\begin{figure}[htb]
\begin{centering}
\includegraphics[width=0.5\textwidth]{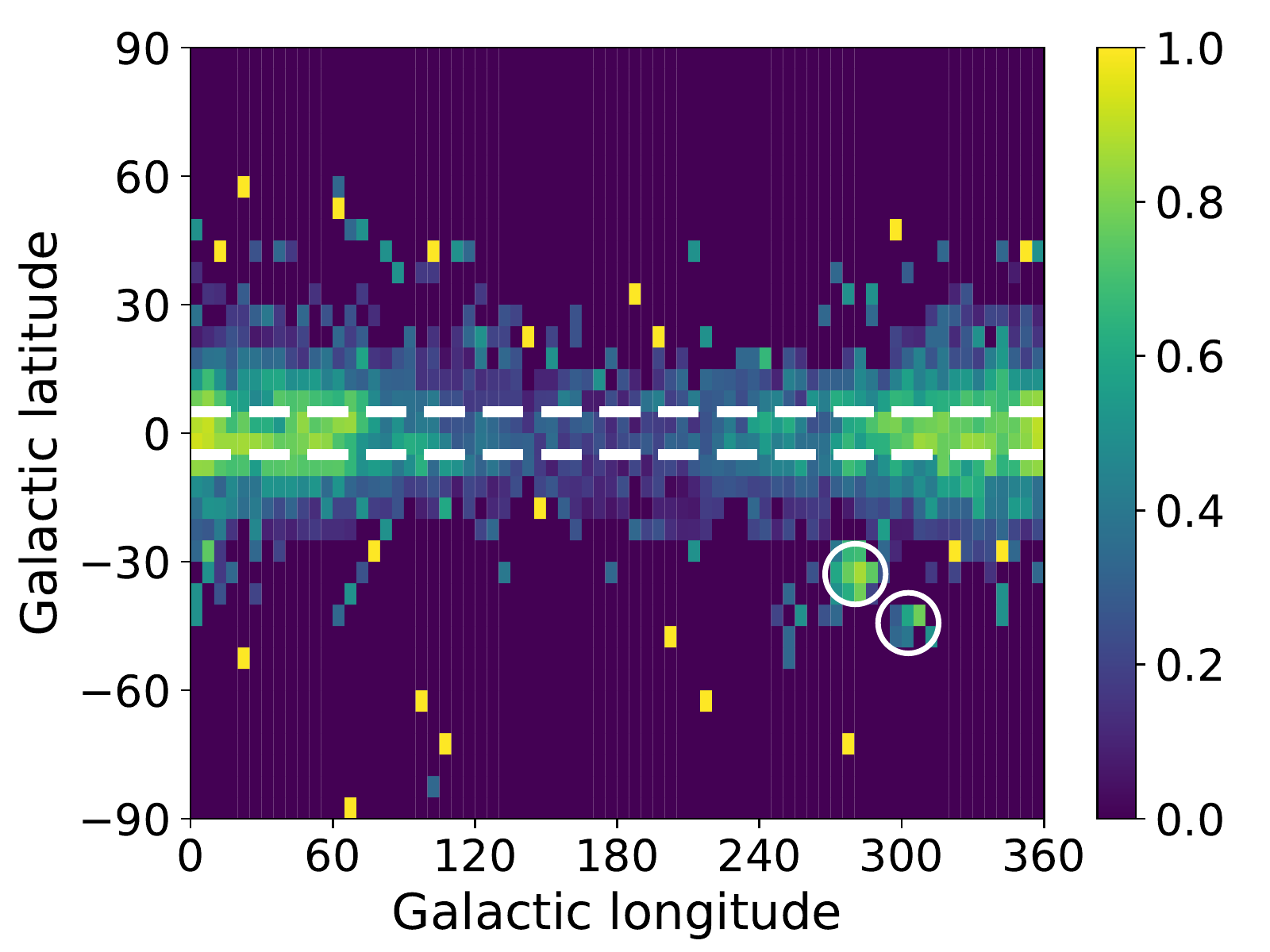}
\caption{Heatmap showing the fraction of complete objects in each $5^{\circ} \times 5^{\circ}$ bin that are ``multiple.'' Redder colors indicate where a larger fraction of the total has been classified as ``multiple.'' Dark blue indicates that a bin contains no multiples. Multiples are more prevalent in the plane (between white dashed lines) of the Galaxy and toward Galactic center. Outside of the Galactic plane, a higher density of ``multiples'' is observed in the Large and Small Magellanic Clouds (locations indicated by the white circles). }
\label{fig:classification_data_latlong}
\end{centering}
\end{figure}

Figure \ref{fig:multiple_rate} shows the rate of multiples as a function of Galactic latitude. ``Multiples'' dominate especially in the Galactic plane; $74.84\%$ ($\pm 0.16\%$) of subjects in the $-5^{\circ} < b < 5^{\circ}$ range were classified as multiples, while only $4.74\%$ ($\pm 0.30\%$) of subjects in that range are good, with over 100,000 complete subjects in this range. We also observe a large spike in ``multiple'' rate between Galactic latitude $-35^{\circ} < b < -30^{\circ}$. Figure \ref{fig:classification_data_latlong} shows the density of ``multiples'' (i.e. the fraction of subjects that are multiples) as a function of Galactic latitude and longitude (redder indicates a larger fraction); this plot reveals that the spike in the $-35^{\circ} < b < -30^{\circ}$ range in Figure \ref{fig:multiple_rate} is associated with the LMC. Figure \ref{fig:classification_data_latlong} also shows a higher density of ``multiples'' in the SMC than in its surroundings. Our statistics are poor in the $60^{\circ} < b < 90^{\circ}$ range because there are fewer complete subjects in this region of sky, which only recently became active on the website.

\subsection{Literature Review by Citizen Scientists}
The next stage of our process is to review the published literature on each of the good objects, discard point-source false positives like known M giants, classical Be stars and AGN, and note re-discoveries of well-studied disk systems. While this task began as a science team function, we have since trained members of the advanced user group to use SIMBAD and VizieR themselves to perform some of this work, as detailed in Paper 1. These citizen scientists also reexamine the website flipbooks for these subjects as an additional check on the website classification. We require a minimum of two opinions on each object before a good candidate can become a Disk Detective Object of Interest (DDOI), an object worthy of additional follow-up observations. We decided that any previously-known, well-studied disk systems (e.g. systems observed by the \textit{Spitzer Space Telescope}, or with coronagraphic disk images) should not be designated as DDOIs, because our observing resources were better used on confirming newly detected excesses rather than following up well-studied disks; however, these systems are still valid disk detections, so we do not eliminate them as false positives.


So far, the advanced user team has reviewed 1465 good subjects. Of these, 1011 have become DDOIs, while an additional 252 subjects have not become DDOIs because they are known well-studied disk systems, for a false positive rate from literature review of $\mathbf{14\%}$.


\section{Follow-up Target Selection and Observations}
\label{sec:target_selection}

We have performed follow-up imaging of a subset of our DDOIs using the Robo-AO on the 1.5-m telescope at Mount Palomar and the RetroCam instrument on the Iren\'ee R. Dupont telescope at Las Campanas Observatory. Here we describe our target selection, observations, and reduction methodology for each telescope.








\subsection{Robo-AO Observations}


We observed 230 targets in the Sloan-$i$ filter \citep{2014ApJ...790L...8B} with adaptive optics using Robo-AO. Dates of observations and number of targets observed on each date are listed in Table \ref{table:RoboAO_obs}. 
Targets were selected for visibility from Mt. Palomar, 2MASS $J < 14.5$ and $i < 17$, the limiting magnitude of the telescope. Targets were observed as a sequence of full-frame-transfer detector readouts at the maximum rate of 8.6 Hz for 90 seconds of total integration. We corrected the individual images for detector bias and flat-fielding effects, and then combined them using post-facto shift-and-add processing, using the target star as the tip-tilt star with 100\% frame selection to synthesize a long-exposure image. Additionally, we synthesized shorter-exposure images by selecting smaller percentages of frames based on quality, as in lucky imaging. In most cases, these yielded an inner working angle for detecting background objects of $\sim 0.15''$, far less than the $6''$ of the $W4$ half-width at half-maximum (HWHM).

\begin{deluxetable}{lc}
\tablecaption{Summary of Observations with Robo-AO \label{table:RoboAO_obs}}
\tablehead{\colhead{UT Date} & \colhead{Objects Observed}}
\startdata
2014-06-14 & 15 \\
2014-06-15 & 12 \\
2014-06-16 & 2 \\
2014-06-19 & 1 \\
2014-07-12 & 1 \\
2014-07-13 & 18 \\
2014-07-14 & 1 \\
2014-07-16 & 1 \\
2014-07-17 & 6 \\
2014-07-18 & 1 \\
2014-08-21 & 2 \\
2014-08-29 & 4 \\
2014-08-30 & 2 \\
2014-08-31 & 65 \\
2014-09-01 & 3 \\
2014-09-02 & 1 \\
2014-09-03 & 1 \\
2014-11-06 & 1 \\
2014-11-09 & 9 \\
2014-11-10 & 26 \\
2014-11-11 & 41 \\
2015-03-08 & 3 \\
2015-03-09 & 18 \\
\enddata
\end{deluxetable}

\subsection{Dupont/RetroCam}

We also collected high-resolution images of 166 targets (including 15 observed with Robo-AO) in the Yc ($\lambda_{c} = 1.035 \mu$m) and Hc ($\lambda_{c} = 1.621 \mu$m) filters using RetroCam on the 100-inch (2.54-m) Irene\'e R. DuPont Telescope at Las Campanas Observatory. Dates of observations and number of targets observed on each date are listed in Table 2. 
Targets were selected for visibility from Las Campanas and primarily for 2MASS $J < 14.5$. We observed targets using a five-point ``dice'' dither pattern. Individual images were corrected for dark current, and flat-fielded using the difference between lamp-on and lamp-off dome flats. $H$-band sky images were generated for each target by median-combining images in a dither sequence without aligning them, removing stellar contributions via sigma-clipping. The $H$-band science images were then sky-subtracted, using the sky image for that target. Once images were completely reduced (dark, flat, and sky corrected), images for each target in each band were then combined with the \texttt{imalign} and \texttt{imcombine} procedures in IRAF, using bright stars in the field of view (including the target star) to align the image stack. Seeing in these observations (as measured by the FWHM of a star other than the target in the stacked image) was generally $\sim 0.8''$, still smaller than the $W4$ HWHM. 


\begin{deluxetable}{lccc}
\tablecaption{Summary of Observations with RetroCam/Dupont \label{table:Dupont_obs}}
\tablehead{\colhead{} & \colhead{Objects} & \colhead{Minimum Total} & \colhead{Maximum Total} \\
\colhead{UT Date} & \colhead{Observed} & \colhead{Integration Time} & \colhead{Integration Time}}
\startdata
2015-06-30 & 4 & 25 & 25 \\
2015-07-01 & 52 & 25 & 100 \\
2015-07-02 & 27 & 25 & 100 \\
2015-10-26 & 19 & 25 & 100 \\
2015-10-27 & 39 & 25 & 100 \\
2015-10-28 & 40 & 25 & 100 \\
\enddata
\end{deluxetable}

\section{Image Analysis}
\label{sec:visual_analysis}







A group of ten citizen scientists examined the Robo-AO data with the SAOImage \texttt{DS9} software package to visually examine the data to identify images with faint background objects, providing them with a set of images analyzed by the science team as a training set. We developed the following method for qualitative analysis:

\begin{enumerate}
\item Display the image with 100\% frame selection as black/white grayscale, min/max, linear scaling, to identify the target star.
\item Shift to ``\texttt{zscale}'' to look for fainter objects. This allows fainter background objects to emerge more clearly than they would otherwise.
\item Identify any faint background objects in the field of view, noting their positions.
\item Note which objects have background objects within 12 arcseconds of the target star.
\end{enumerate}

Each volunteer independently analyzed a subset of the images using the above method. The group then discussed each image together to reach a consensus on each target. We followed a similar procedure with the Dupont data. 

This visual inspection identified targets with evidence of background sources that could produce a false positive excess. We then quantified whether these background sources significantly affected the excess at W4. To estimate the contribution each background object made at W4, we determined magnitudes for the background objects with aperture photometry using the IRAF \texttt{DAOphot} package. We assumed that background objects exhibited either an M dwarf SED, or a power-law SED with spectral index 0 (corresponding to a YSO or heavily-reddened early-type star), and determined the colors of these objects (calculated in Appendix \ref{sec:appendix} and presented in Table \ref{table:colors}). Using the recorded photometry and these colors, we estimated the flux of the background object at W4, and subtracted this flux from the total W4 flux to yield the intrinsic flux of the target itself (including any contribution from circumstellar material). We then re-calculated the target's [W1-W4] color using this corrected W4 flux to determine if a significant excess remained. The results are described in Section \ref{sec:multiple_rates}.

\section{False Positive Rates of AllWISE Disk Detections}
\label{sec:multiple_rates}



In this section, we determine false-positive rates based on the follow-up observations and unresolvable IR-bright background objects. We then combine these with the classification and literature-review data to determine the overall false-positive rate for Disk Detective thus far, from which we estimate the eventual final yield of disks from AllWISE.

\subsection{False-Positive Rates from High-Resolution Follow-up}


The results of quantitative analysis of the 261 targets with reliable photometry observed with Palomar/Robo-AO and Dupont/Retrocam are presented in figure \ref{fig:followup}. We combine the two samples without adjustment because (as described in Appendix \ref{sec:appendix}) both instruments are similarly sensitive to the same background objects. Overall, 244 of these 261 targets retain a significant infrared excess after the contribution of background objects has been removed, for a false-positive rate of $7\% \pm 1\%$. There is no detectable significant difference in contamination rate in the Galactic plane due to the relatively small numbers involved (compared to the overall Disk Detective input catalog). Of the 39 objects in the $-5^{\circ} < b < 5^{\circ}$ range, 3 are contaminated, leaving a false positive rate of $8\% (\pm 4\%)$. Out of the plane, 14 of 222 targets are contaminated, leaving a false positive rate of $6\% (\pm 2\%)$.


\begin{figure}[htb]
\begin{centering}
\includegraphics[width=0.5\textwidth]{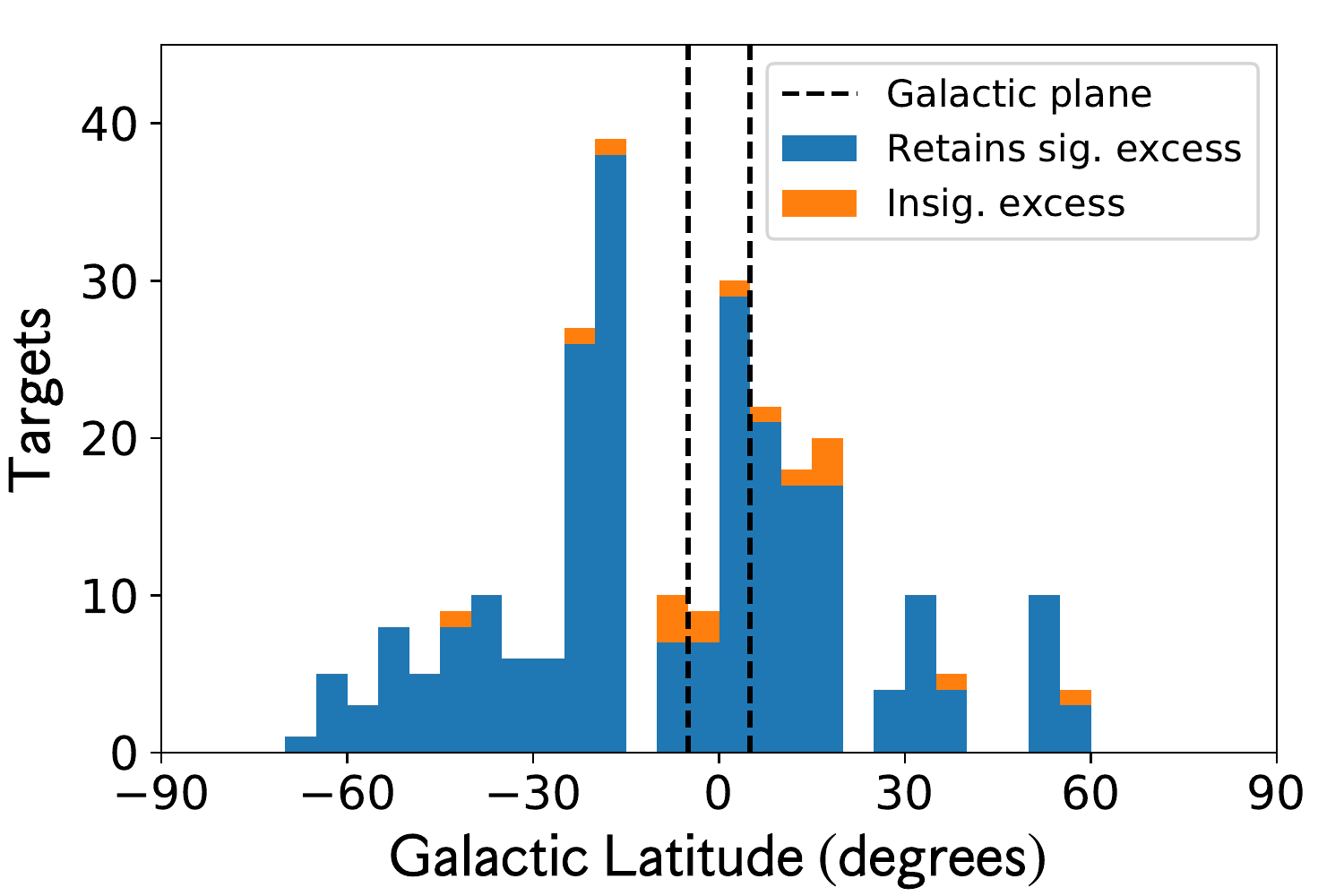}
\caption{Objects with high-resolution imaging from Robo-AO and Dupont as a function of Galactic latitude. The overall false-positive rates are low in comparison to the rates from our website-based analysis. The rates inside and outside the Galactic plane show no significant difference.}
\label{fig:followup}
\end{centering}
\end{figure}

We identified 16 objects as having insufficient excess at W4 once the estimated contributions from background objects were removed. We list these objects in Table \ref{table:bad_objects}, as well as any previous identifications as excess targets.

\begin{deluxetable}{ccl}
\tablecaption{Targets with False-Positive Excesses Due to Background Objects in High-Resolution Follow-up Observations \label{table:bad_objects}}
\tablehead{\multicolumn{2}{c}{Identifiers} & \colhead{Previous} \\ \colhead{Zooniverse} & \colhead{WISEA} & \colhead{Citations}}
\startdata
AWI0005w52 & J000308.37+424452.4 & \\
AWI0005yiz & J010722.60+380143.9 & 1 \\
AWI0002vbd & J020206.67+601741.4 & \\
AWI0003cs7 & J021532.17+591424.4 & \\
AWI0004nfu & J032853.67+490412.8 & 1 \\
AWI0000uj2 & J153046.05+342756.4 & 1,2,3 \\
AWI00006nb & J161808.08+104551.4 & \\
AWI0005bk2 & J172912.43+005605.7 & \\
AWI0005brq & J181949.03+310841.7 & \\
AWI0005bud & J184141.31+313703.4 & \\
AWI0005lx9 & J193040.05+350609.9 & \\
AWI00055by & J204443.79+425654.9 & 4 \\
AWI0005vyx & J212959.78+413037.3 & \\
AWI0005a9r & J215305.45+682955.0 & 1 \\
AWI0005w1h & J220503.97+444543.7 & \\
AWI0000kk6 & J220601.14-020343.2 & \\
\enddata
\tablerefs{(1) Paper 1. (2) \citet{2013ApJS..208...29W}. (3) \citet{2016ApJS..225...15C}. (4) \citet{2005MNRAS.363.1111C}}
\end{deluxetable}



\subsection{False-Positive Excess from Unresolvable Infrared Galaxies}
\label{sec:ulirgs}

While the Robo-AO and RetroCam observations catch many otherwise-unresolved background sources, they leave one potential source of confusion unexplored: objects clustered on the scale of the W4 beam with no counterpart in near-IR or red-optical light, such as luminous or ultra-luminous infrared background galaxies, or (U)LIRGS \citep{Papovich2004}. While WISE is not well-suited to exploring the density of these phenomena, previous higher-resolution mid-infrared surveys, operating at similar wavelengths, can provide constraints. \citet{Papovich2004} used \textit{Spitzer}/MIPS data to identify a previously-undetected population of infrared-luminous galaxies, quoting a cumulative distribution of number of galaxies as a function of source brightness at 24 $\mu$m. We can use this distribution to estimate the confusion noise from these galaxies in our 22 $\mu$m data if we correct for sources that would be detected in website classifications and follow-up imaging.


We determined the minimum flux at W4 for a background galaxy to produce a false positive excess, assuming a bare stellar photosphere in the Rayleigh-Jeans tail. We also estimated the flux at W4 at which a background galaxy would produce a visible signal in 2MASS $H$ images, assuming that such an object would be visible if its 2MASS $H$ flux were half that of the target. We applied these numbers to every good subject, finding the cumulative number of sources per steradian bright enough to produce a false positive, and subtracting from that the cumulative number of sources per steradian bright enough to have been detected as a background object in shorter-wavelength data. We then multiplied by the angular beam width at W4 to get the expected number of sources. We treated this as a probability of the number of sources in the beam for each object, summing the probability for each object to estimate the number of good objects for which a background galaxy was the source of the false-positive.


Of the 14,681 complete good subjects, we expect that $7.4 \pm 2.7$ subjects, or $0.05\% \pm 0.02\%$, are contaminated by an unresolved background infrared-luminous galaxy, a negligible contribution.




\subsection{Expected Total Number of Disks in AllWISE}

Combining the data from user classifications, advanced user review, high-resolution follow-up observations, and background galaxy count estimates, we find that of the 149,273 complete subjects on the Disk Detective web site, $7.9\%$ ($\pm 0.2\%$) are likely to be stars with circumstellar material. $90.20\%$ of subjects are eliminated by website evaluation; $1.35\%$ were eliminated by literature review; $0.52\%$ were eliminated by follow-up observations; and a near-negligible fraction ($<0.01\%$) are expected to have a false-positive due to undetectable background galaxies.

Applying this number to the full brightness-limited sample currently on the website, we would expect to find $\sim\,21,600$ disk candidates overall, out of 272,022 subjects in the brightness-limited input catalog. Given that this analysis doesn't incorporate false positives that are only identifiable in spectroscopic follow-up (e.g. M giants, classical Be stars), we expect that this number is an upper limit to the number of debris and YSO disks with W4 excess in the AllWISE catalog. This 8.0\% figure is higher than the 3.4\% found by \citet{2012MNRAS.426...91K}. We attribute some of this difference to our full-sky scope, as \citet{2012MNRAS.426...91K} only considered sources in or near the Galactic plane. We also hypothesize that some of the difference can be attributed to our detailed visual inspection of each source, rather than adopting a simple across-the-board cut based on 100$\mu$m flux, as they did. 



\section{Application to Other WISE Disk Searches}
\label{sec:other_surveys}

Given its large input catalog compared to other surveys, Disk Detective is well positioned to not only identify new warm debris disks in the WISE catalog, but inspect and re-evaluate disks identified by other researchers. In this section, we estimate the numbers of published disks from other searches that are likely false positives, and present a visual re-inspection of the M dwarf disk sample presented by \citet{2014ApJ...794..146T}.


\subsection{False-Positives in Previous WISE Disk Searches}

\begin{figure}[htb]
\begin{centering}
\includegraphics[width=0.5\textwidth]{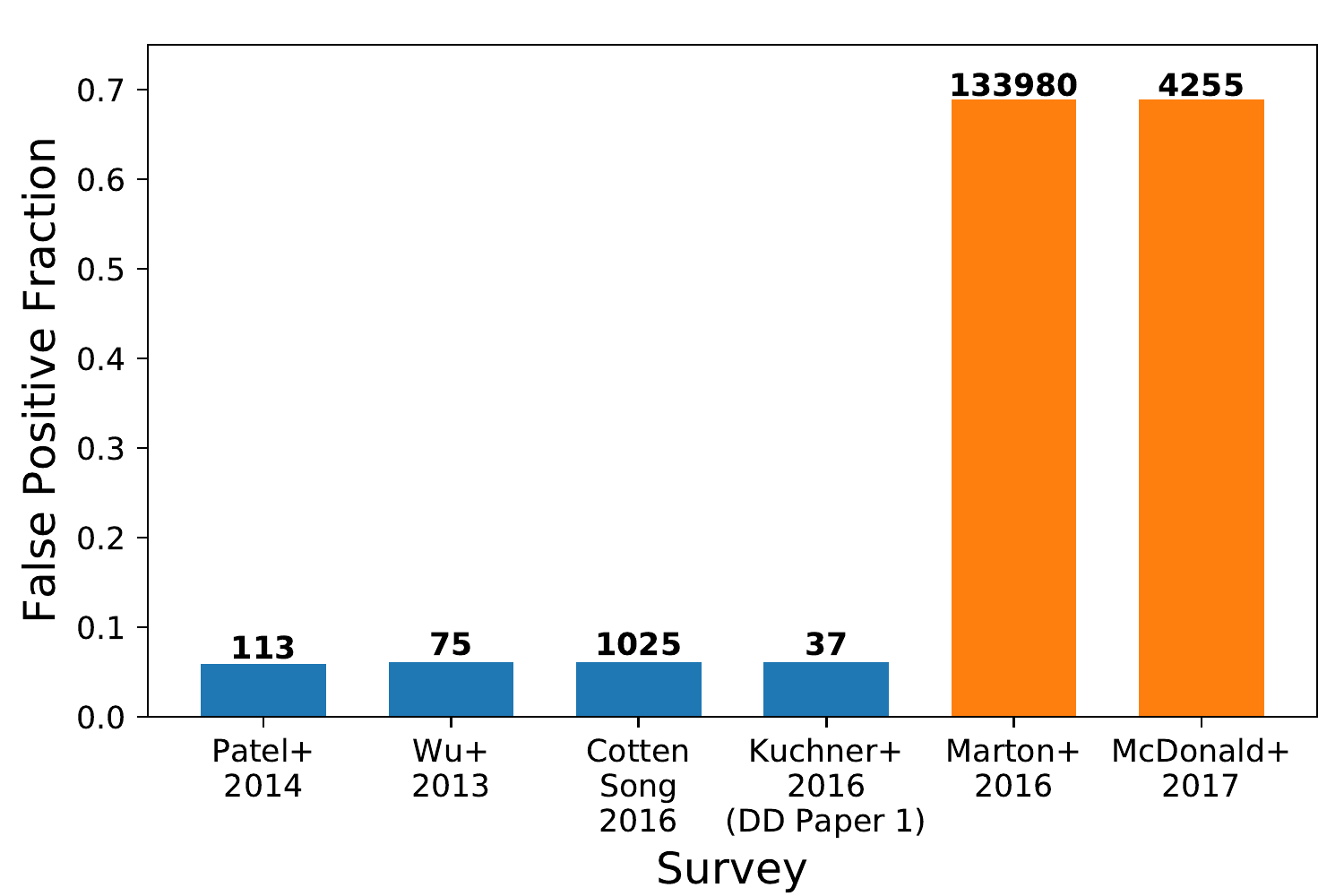}
\caption{Expected false-positive fractions for published WISE disk searches. Searches that incorporate visual inspection of images (blue bars) have lower expected false-positive fractions than those that do no incorporate visual inspection (orange bars). Numbers above each bar indicate the number of disk candidates in each search.}
\label{fig:false_positive_fraction}
\end{centering}
\end{figure}

Because our sample encompasses the full 2MASS/WISE cross-match, we can apply our false positive rates to other searches for warm debris disks with WISE. While we cannot assess individual objects in other surveys due to limited overlap with our follow-up, we estimate the number of disks in each survey likely to be a false positive based on the rates we determined, depicted in Figure \ref{fig:false_positive_fraction}.

\begin{itemize}
\item \textbf{\citet{2014ApJS..212...10P}.} This survey specifically avoided the Galactic plane in its cross-match with the \textit{Hipparcos} catalog \citep{2007A&A...474..653V}, so we similarly apply an out-of-plane contamination rate. We will assume that the visual inspection described by the authors is comparable to the inspection and literature review by the Disk Detective team, and will thus only apply the rate of contamination from follow-up imaging. Of the 113 new warm disks that were detected in this survey \citep{2015ApJS..220...21P}, we expect $\sim 7$ to have previously-undetected background objects that would contribute to a false-positive excess detection. 


\item \textbf{\citet{2013ApJS..208...29W}.} This survey did not specifically avoid the Galactic plane as a selection criterion in its cross-match with \textit{Hipparcos}, so we apply the statistics for the full sample to this paper's sample of 75 previously-unidentified main sequence stars with infrared excess indicative of a disk. We would therefore expect $\sim 5$ of these targets to be contaminated by one or more otherwise-undetected background objects. 

\item \textbf{\citet{2016ApJS..225...15C}.} This search identified $\sim 1750$ debris disk candidates from a thorough review of pre-WISE disk literature and a cross-match of AllWISE with the Tycho-2 \citep{2000A&A...355L..27H} survey, of which 1025 are new excess detections. Applying our recovery rate for high-resolution follow-up to this sample, we would expect that $\sim 64$ targets would be contaminated by background objects only recoverable in high-resolution images that could produce a false-positive excess detection.

\item \textbf{\citet{2012MNRAS.427..343M}, \citet{2017MNRAS.471..770M}.} These searches cross-matched photometry data from \textit{Hipparcos}, \textit{Tycho}, and various other catalogs, using astrometry from, respectively, \textit{Tycho} and \textit{Gaia}. These works then used SED fitting of the cross-matched data to estimate stellar and excess parameters for $>100,000$ objects. However, neither search considered the source of the excess (e.g. self-produced dust around AGB stars rather than debris disks), and neither search included a robust visual inspection of the actual images (which eliminates $\approx 90\%$ of excess candidates in the Disk Detective search), relying instead on visual inspection of the fit SEDs. Because these searches select for objects with star-like SEDs rather than preselecting sources with infrared excess the way our search does, we cannot directly apply our overall false-positive rates to their search. However, since we estimate there to be $\sim 21,600$ disk candidates in AllWISE, we expect that $>79,000$ of the excesses identified by \citet{2012MNRAS.427..343M} and \citet{2017MNRAS.471..770M} are false positives.

\item \textbf{\citet{2016MNRAS.458.3479M}.} This search used support vector machines (SVM), a class of supervised learning algorithm, to identify 133,980 YSO candidates in AllWISE, by identifying objects that were a given type of false positive in the data via SIMBAD, and using these to train the SVM to avoid such objects. However, there was no visual inspection of the images, sources were not pre-selected for excess, the W4 PSF was not taken into account, and there is no acknowledgement of multiples, which make up the bulk of our false-positives. If we assume that the SVM algorithm identifies non-multiple false positives as effectively as Disk Detective, literature review would eliminate the SIMBAD-identified non-YSOs in their final sample (as per Table 1 of that paper), and also assume a similar number of objects would be false positives in follow-up imaging, this leaves a lower limit false-positive rate of 74\%. While a search through WISE using machine-learning algorithms could prove valuable, the number of false positives that are only identified by visual inspection of images suggests that any such system would need to take the images themselves into account, rather than strictly learning based on photometric points and data quality flags.

\item \textbf{\citet{Kuchner2016}.} As Paper 1 was published before analysis of Robo-AO and Dupont photometry was complete, we also apply our rate analysis to this paper. Of the 37 new disk candidates presented in that paper, we expect that $\sim 2$ would be contaminated.



\end{itemize}



The targets WISEA J153046.05+342756.4 (AWI0000uj2), WISEA J010722.60+380143.9 (AWI0005yiz), WISEA J032853.67+490412.8 (AWI0004nfu), and WISEA J215305.45+682955.0 (AWI0005a9r) present salient examples of the importance of imaging follow-up. AWI0000uj2, an A0 star, appears in \citet{2013ApJS..208...29W}, in \citet{2016ApJS..225...15C} as a ``Reserved'' excess candidate, \textit{and} in Paper 1. AWI0004nfu appears in Paper 1 and \citet{2012ApJ...752...58Z}. AWI0005yiz and AWI0005a9r were both detected for the first time in Paper 1.

Based on our follow-up observations with Robo-AO, all four of these objects have no excess once the contribution from background objects is subtracted at W4. AWI0000uj2 is likely significantly contaminated at W4 by two objects $\sim 4$ magnitudes fainter than it in the Sloan $i$ band. AWI0004nfu exhibits seven background objects $3-9$ magnitudes fainter than it. AWI0005yiz exhibits several background objects $\sim 5-8$ magnitudes fainter than it, and AWI0005a9r exhibits several background objects $3-10$ magnitudes fainter than it in Sloan $i$. Further analysis and observations (e.g. additional wavelength coverage of the background objects to confirm the shape of their SED) are required to confirm that the observed excess is significantly affected by these targets. However, these cases illustrate that high-resolution follow-up can impact the quality of published infrared excesses.

\subsection{A Re-Assessment of a Previous WISE-based M Dwarf Disk Search}


M dwarf disk systems are particularly invaluable systems to identify. M dwarfs are key targets for large-scale exoplanet searches \citep[e.g.][]{2009IAUS..253...37I, 2014SPIE.9143E..20R}, and host some of the most unique exoplanetary systems discovered to date \citep[e.g.][]{2016Natur.536..437A, 2017Natur.542..456G}. Debris disks around M dwarfs should have the same informative powers as their higher-mass-star cousins. The recent discovery of cold debris around the M dwarf exoplanet host Proxima Centauri with ALMA \citep{2017ApJ...850L...6A} may yield valuable insights into the structure of that planetary system, though a recent re-analysis indicates that much of the flux previously attributed to dust can be attributed to a large flare \citep{2018ApJ...855L...2M}.

However, the current sample of M dwarf debris disk systems has significant shortcomings. Only $\sim 2\%$ of dM stars are currently known to host debris disks \citep[e.g.][]{2013MNRAS.434.1117D, 2014ApJ...794..146T, 2017MNRAS.469..579B, 2016ApJ...832...50B, 2016ApJ...830L..28S}, less than the $\sim 14\%$ expected from the mass distribution of primordial disks, and much less than the $20\%$ fraction of A dwarfs with known disks \citep{2010MNRAS.409L..44G}. As part of our application of Disk Detective analysis to other surveys, we began a re-analysis of imagery of the M dwarf disk candidates presented in \citet{2014ApJ...794..146T}, one of the largest lists of M dwarf disk candidates from WISE published to date. 

These disk candidates generally suffer from low signal-to-noise at W4. Of the 175 disk candidates presented in \citet{2014ApJ...794..146T}, only three meet the Disk Detective input catalog criterion of $\texttt{w4snr} > 10$. By contrast, 152 of the 175 only have upper limit magnitudes at W4 due to $\texttt{w4snr} < 2$, as per the AllWISE data release quality flags. In most cases, W4 postage stamp images only show background emission, without a coherent point source at W4. This is presumably due in combination to the shallowness of the W4 band, and the distance of these targets.

Because of this lack of a point source, the Disk Detective classification method fails for the majority of these targets. Due to the shallowness of W4 compared to the other three WISE bands, \citet{2014ApJ...794..146T} primarily focused on significant excess at W3, noting that those targets with $\texttt{w4snr} > 3$ and excess at W4 also exhibited excess at W3. However, the Disk Detective method is still viable for the 13 targets with $\texttt{w4snr} > 3$. 

We downloaded 1-arcminute postage-stamp images of the \citet{2014ApJ...794..146T} disk candidates from the IRSA finder chart, using a similar blue-white color scale to that used on the Disk Detective website, and applied a 12-arcsecond radius red circle to these images to effectively generate Disk Detective flipbooks of these targets. Our team of advanced users then analyzed these targets as if they appeared on the Disk Detective website and cataloged their assessment of these objects in the Disk Detective website categories. All thirteen objects with $\texttt{w4snr} > 3$ were classified as extended by a majority of classifiers. Five of the thirteen were also majority-classified as multiple, and two were also majority-classified as oval.

The disk candidates identified by the \citet{2014ApJ...794..146T} M dwarf survey typically do not meet the Disk Detective standard for inclusion for analysis; the thirteen that meet these standards are clear false positives that do not survive Disk Detective's by-eye examination. It is possible that the remaining 162 M dwarf disk candidates, with significant W3 excess and low signal-to-noise at W4, could represent a class of hot, distant debris disks for which Disk Detective is not designed. However, given the high false positive rate of objects that can be assessed with the Disk Detective methodology, we recommend treating these results with caution. 

\section{New Disk Candidates}
\label{sec:new_candidates}

\begin{deluxetable*}{lllccccccl}
\tablecaption{Disk Candidates \label{table:candidates}}
\tablehead{\multicolumn{3}{c}{Identifiers} & \colhead{Sp.} & \colhead{Distance\tablenotemark{b}} & \multicolumn{4}{c}{Magnitudes} & \\ \colhead{Zooniverse} & \colhead{HD} & \colhead{WISEA} & \colhead{Type\tablenotemark{a}} & \colhead{(pc)} & \colhead{Vmag\tablenotemark{c}} & \colhead{Jmag} & \colhead{W1mag} & \colhead{W4mag} & \colhead{Notes}}
\startdata
AWI0000gib & 1777   & J002133.47-661816.6 & A0V       & $163 \pm 10$  & $ 7.361 \pm 0.010$ & $ 7.365 \pm 0.027$ & $ 7.356 \pm 0.027$ & $6.793 \pm 0.066$ & \\
AWI0006251 &        & J002155.14-672715.9 &           & $789 \pm 231$ & $10.75  \pm 0.06 $ & $ 9.008 \pm 0.023$ & $ 8.253 \pm 0.022$ & $6.655 \pm 0.065$ & \\
AWI00062lo & 2830   & J003140.76-014737.3 & A0V       & $109 \pm 6$   & $ 7.07  \pm 0.01 $ & $ 6.917 \pm 0.019$ & $ 6.841 \pm 0.033$ & $5.832 \pm 0.049$ & a,b \\
AWI0005mry & 3051   & J003412.66+540359.0 & A1V       & $213 \pm 15$  & $ 7.595 \pm 0.010$ & $ 7.350 \pm 0.019$ & $ 7.346 \pm 0.049$ & $6.881 \pm 0.067$ & a \\
AWI0000jvv &        & J003507.14+070625.0 &           &               &                    & $ 8.142 \pm 0.019$ & $ 6.937 \pm 0.033$ & $4.923 \pm 0.035$ & c \\
AWI00062m4 &        & J004826.42+020753.0 &           & $535 \pm 213$ & $10.264 \pm 0.046$ & $ 8.490 \pm 0.026$ & $ 7.687 \pm 0.024$ & $5.392 \pm 0.043$ & c \\
AWI0005yiv & 5741   & J005926.26+400918.2 &           & $113 \pm 4$ & $ 7.532 \pm 0.010$ & $ 7.170 \pm 0.018$ & $ 7.139 \pm 0.031$ & $6.651 \pm 0.054$ & \\
AWI00055sx & 6370   & J010652.55+743754.5 & B9IV      & $354 \pm 39$ & $ 8.368 \pm 0.012$ & $ 8.168 \pm 0.029$ & $ 8.149 \pm 0.023$ & $7.139 \pm 0.087$ & a \\
\enddata
\tablenotetext{a}{Spectral types are from SIMBAD, with the exception of sources that appear in Paper 1. For those sources, we use the spectral types published in that paper.}
\tablenotetext{b}{Distances are based on parallax measurements from \textit{Hipparcos} or TGAS, as listed on SIMBAD.}
\tablenotetext{c}{Sourced from SIMBAD.}
\tablecomments{(a) Appears in Paper 1. (b) Listed as a comoving object in \citet{2017AJ....153..257O}. (c) Listed in the K2 Ecliptic Plane Input Catalog. (d) Listed in \citet{2016NewA...44....1C}. (e) Identified in \citet{2003AJ....126.2971V}. (f) Identified in \citet{2009BaltA..18....1C}. (g) Identified in \citet{2016ApJS..225...15C}. (h) Identified in \citet{2014ApJ...784..126E}. (i) Identified in \citet{2011ApJS..196....4R}. (j) Identified in \citet{2013ApJS..208...29W}. (k) Identified in \citet{2010ApJ...720...46G}. (l) Identified in \citet{2015AJ....150..100K}. (m) Identified in \citet{2005AJ....129..856H}. (n) Identified in \citet{1996AAS...119....7A}. (o) Identified in \citet{2008MNRAS.387.1335R}. (p) Identified in \citet{2014AAp...572A..89S}. (q) Identified in \citet{2000AJ....119.3026R}. (r) Identified in \citet{2012AJ....144..192M}. (s) Identified in \citet{2008ApJ...688..362L}. (t) Identified in \citet{2014ApJ...794..124R}. (u) Identified in \citet{2012MNRAS.421L..97R}. (v) Identified in \citet{2012ApJ...756..133C}. (w) Identified in \citet{2014ApJS..211...25C}. (x) Identified in \citet{2013ApJ...778...12M}. (y) Identified in \citet{2009ApJS..181..321E}. (z) Identified in \citet{2003PASP..115..965E}. ($\alpha$) Identified in \citet{2014MNRAS.437..391C}. ($\beta$) Identified in \citet{2005MNRAS.363.1111C}. ($\gamma$) Identified in \citet{2013MNRAS.433.2334K}. ($\delta$) Identified in \citet{2014ApJS..212...10P}. ($\epsilon$) Identified in \citet{2011MNRAS.415..103B}.  \\
Table \ref{table:candidates} is published in its entirety in the machine-readable format. A portion is shown here for guidance regarding its form and content.}
\end{deluxetable*}

Based on our follow-up observations with Palomar/Robo-AO and Dupont/Retrocam, we find that 244 of our observed targets, including 214 sources first identified by Disk Detective, have no significantly contaminating background objects within the $12''$ radius of the W4 PSF, giving us the confidence to publish them as disk candidates. The candidates are listed in Table \ref{table:candidates}; previous surveys that have also identified these targets are listed in the Notes column. We briefly summarize characteristics of some objects of particular interest in Appendix \ref{sec:appendix_candidates}. 

\begin{figure*}[htb]
\begin{centering}
\plottwo{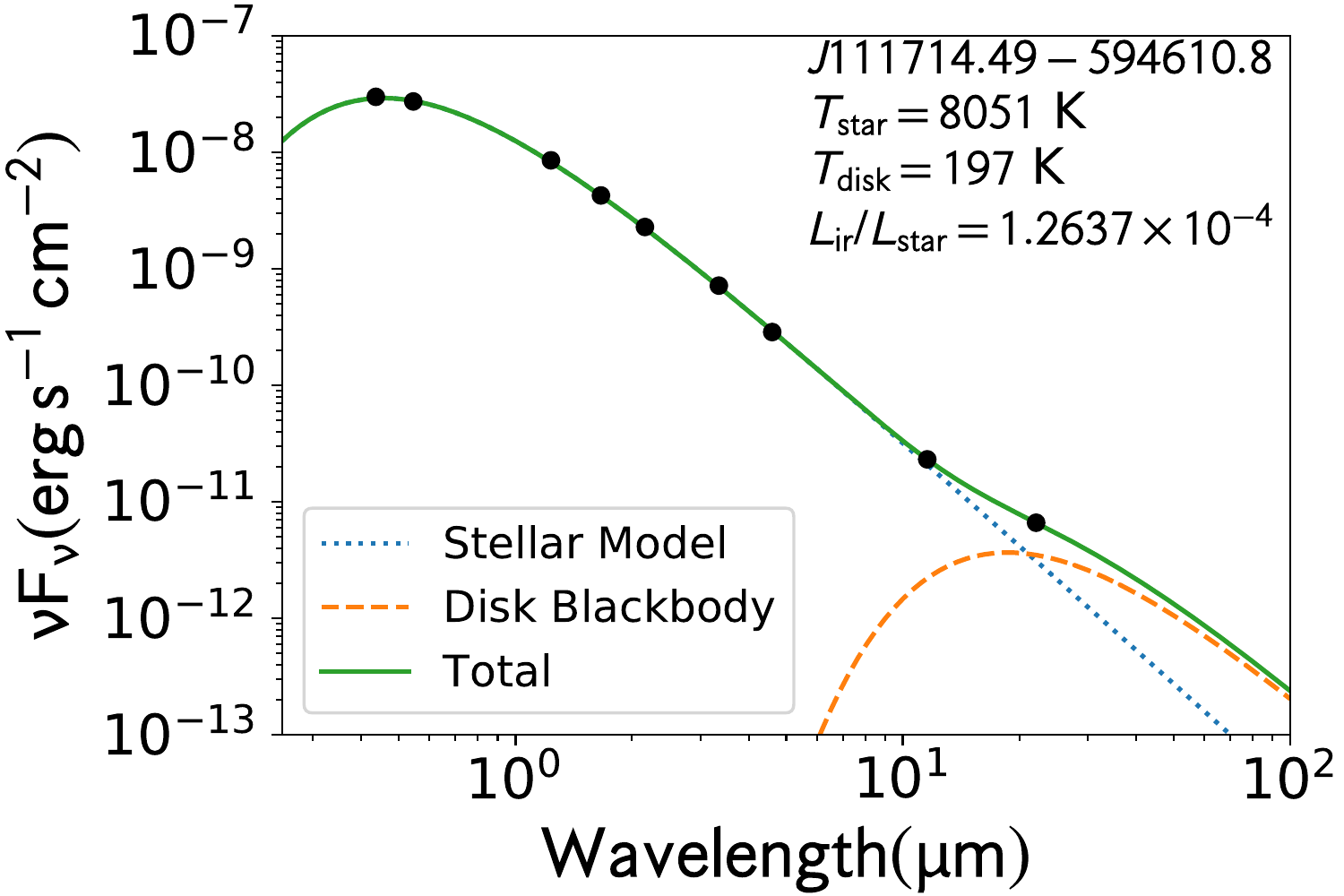}{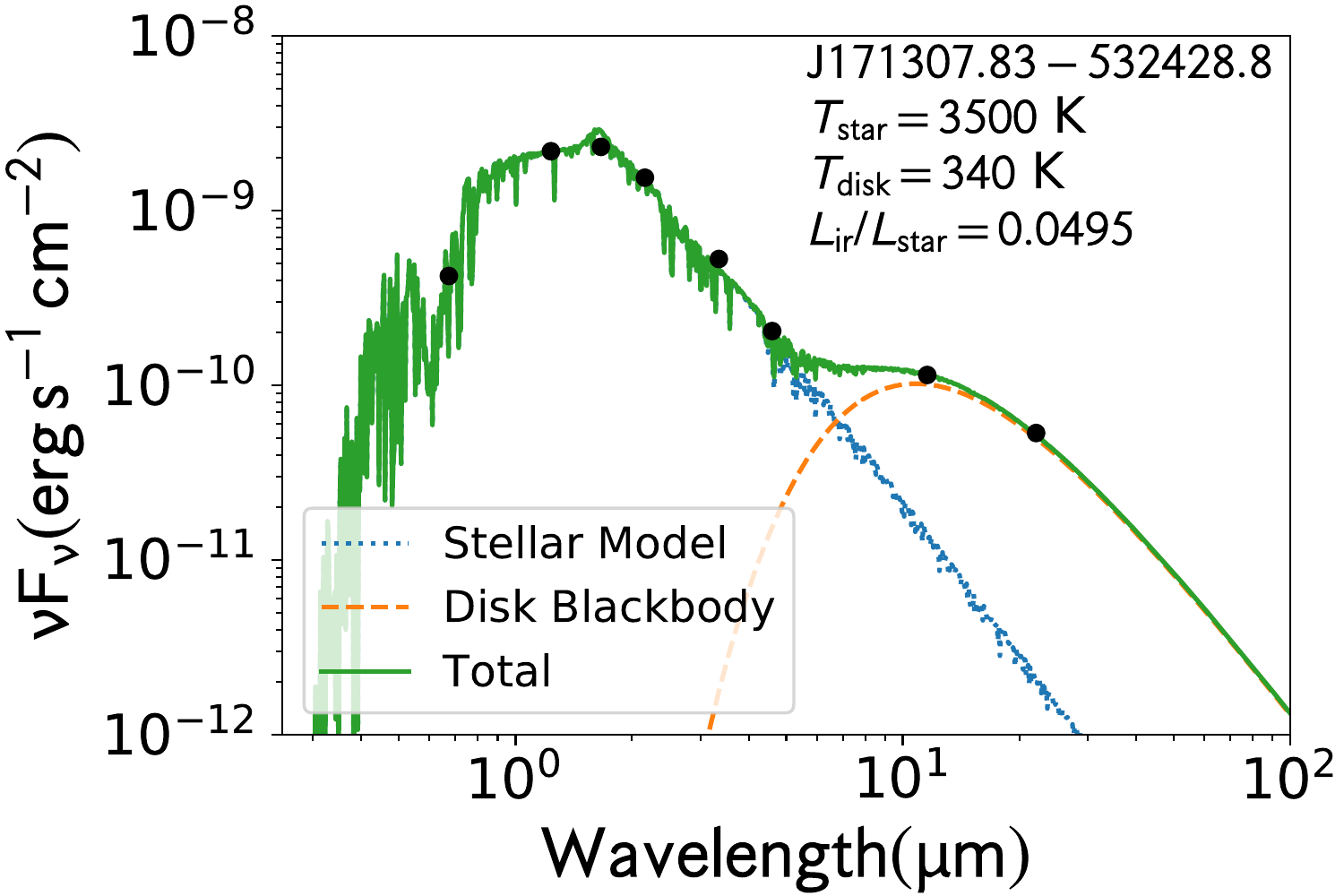}
\caption{Sample SEDs for two of the 244 systems presented here. Plotted are Johnson B and V magnitudes based on \textit{Tycho} photometry (sourced from SIMBAD), 2MASS, and WISE (black circles). Uncertainties in these measurements are typically smaller than the plotting symbol. The best-fit combined model is plotted in green along with each of the model components (photosphere in blue dots, single-temperature dust blackbody as dashed orange). Model stellar temperature, blackbody temperature, and fractional infrared luminosity are listed in the top right corner of each panel. SEDs for the entire sample are available in the online version of the article.}
\label{fig:SED}
\end{centering}
\end{figure*}

We used available photometry from SIMBAD, 2MASS, and WISE (including corrected W1, W2, and W4 photometry) to fit the spectral energy distributions (SEDs) of these systems. We initially fit the stellar component of the system with a blackbody to estimate the stellar temperature and ratio of stellar radius to distance. In the case where a blackbody fit yields a temperature less than 7000K, we instead fit with a stellar model from the BT-Settl CIFIST package \citep{2015A&A...577A..42B}, also fitting for $\log(g)$. We initially fit the three bluest photometry points, then refit including the next bluest point if it is not in significant excess, repeating the last step iteratively until the next point is either W3 or in excess. We fit the remaining excess using a single-temperature blackbody to determine the dust temperature. We then find the best fit parameters using the \texttt{emcee} package \citep{2013ascl.soft03002F}, using the previous parameter estimates as an initial guess. These SEDs, shown in Figure \ref{fig:SED}, provide a useful first-order estimate of these fundamental disk parameters. We list these parameters for likely debris disks in Table \ref{table:debris_disk_parameters}, and for likely YSO disks in Table \ref{table:yso_parameters}. Table \ref{table:debris_disk_parameters} consists of all candidates with only one point of excess (W4), as well as candidates with two points of excess and $L_{\mathrm{ir}}/L_{\star} < 10^{-3}$. Table \ref{table:yso_parameters} consists of all other objects. For objects with more than two points of excess, we also provide the spectral index $\alpha$ of a power-law fit to the object's WISE data as an estimate of YSO class.

\begin{deluxetable*}{llccc}
\tablecaption{Derived Parameters of Debris Disk Candidates \label{table:debris_disk_parameters}}
\tablehead{\colhead{Zooniverse} & \colhead{WISEA} & \colhead{$T_{\mathrm{eff}}$ (K)} & \colhead{$T_{\mathrm{disk}}$ (K)} & \colhead{$L_{\mathrm{ir}}/L_{\mathrm{star}}$}}
\startdata
AWI00062h7 & J002133.47-661816.6 & $11697_{-113}^{+112}$ & $216_{- 24}^{+ 27}$ & $(2.8 \pm 0.4)       \times 10^{-5}$ \\
AWI0000bs0 & J003140.76-014737.3 & $10319_{- 69}^{+ 64}$ & $202 \pm 11       $ & $(8.6 \pm 0.6)       \times 10^{-5}$ \\
AWI0005mry & J003412.66+540359.0 & $ 9680 \pm 72       $ & $ > 35            $ & $< 0.12                            $ \\
AWI0005yiv & J005926.26+400918.2 & $ 9165_{- 52}^{+ 58}$ & $ > 46            $ & $< 0.005                           $ \\
AWI00055sx & J010652.55+743754.5 & $10106_{- 85}^{+ 80}$ & $182_{- 15}^{+ 16}$ & $(8.1 \pm 1.1)       \times 10^{-5}$ \\
AWI00055sz & J011636.23+740136.6 & $ 9011_{- 89}^{+ 99}$ & $685_{-113}^{+125}$ & $(4.5_{-1.3}^{+2.0}) \times 10^{-4}$ \\
\enddata
\tablecomments{Table \ref{table:debris_disk_parameters} is published in its entirety in the machine-readable format. A portion is shown here for guidance regarding its form and content.}
\end{deluxetable*}

\begin{deluxetable*}{llcccc}
\tablecaption{Derived Parameters of YSO Disk Candidates \label{table:yso_parameters}}
\tablehead{\colhead{Zooniverse} & \colhead{WISEA} & \colhead{$T_{\mathrm{eff}}$ (K)} & \colhead{$\alpha_{\mathrm{disk}}$} & \colhead{$T_{\mathrm{disk}}$ (K)} & \colhead{$L_{\mathrm{ir}}/L_{\mathrm{star}}$}}
\startdata
AWI0000nfp & J002155.14-672715.9 & $ 4700 \pm 100      $ & $                     $ & $   509 \pm         21 $ & $(7.7   \pm        0.05  ) \times 10^{-3}$ \\
AWI0000jvv & J003507.14+070625.0 & $ 3500 \pm 100      $ & $                     $ & $   368_{-  11}^{+  12}$ & $(1.23 _{-0.04 }^{+0.05 }) \times 10^{-2}$ \\
AWI00062m4 & J004826.42+020753.0 & $ 4600 \pm 100      $ & $                     $ & $   207_{-   4}^{+   5}$ & $(3.7  _{-0.1  }^{+0.1  }) \times 10^{-3}$ \\
AWI00062mq & J011743.47-523330.8 & $ 6400 \pm 100      $ & $-0.12 \pm 0.01       $ & $   436 \pm          4 $ & $(9.0   \pm        0.2   ) \times 10^{-2}$ \\
AWI0005aeg & J013833.77+780834.3 & $ 6700 \pm 100      $ & $-2.11 \pm 0.02       $ & $  1210_{-  25}^{+  26}$ & $(2.5  _{-0.1  }^{+0.2  }) \times 10^{-2}$ \\
\enddata
\tablecomments{Table \ref{table:yso_parameters} is published in its entirety in the machine-readable format. A portion is shown here for guidance regarding its form and content.}
\end{deluxetable*}

\begin{figure}[htb]
\begin{centering}
\includegraphics[width=0.5\textwidth]{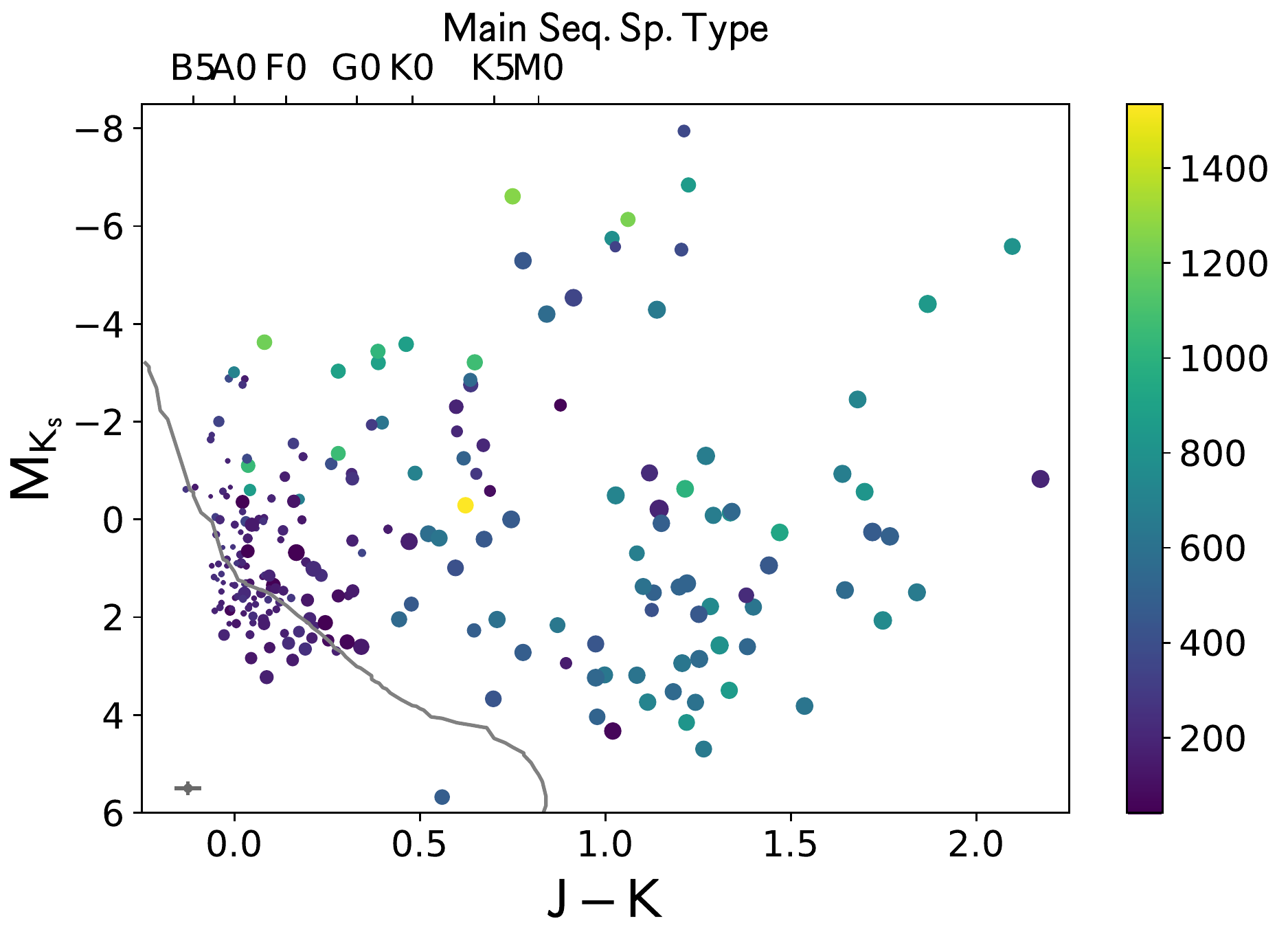}
\caption{2MASS $(M_{K}, J-K)$ color-magnitude diagram of 223 disk candidates with parallax measurements. Point color corresponds to disk temperature, while point size corresponds to $L_{\mathrm{ir}}/L_{\mathrm{star}}$. 148 disk candidates lie within 1.5 mag of the zero age main sequence (gray line), and 126 lie within 1.5 mag of the zero age main sequence with spectral type earlier than G0. Main sequence spectral types are listed across the top for reference. The point in grey in the lower left indicates typical uncertainties, $\sim0.04$ mags in (J-K) color and $\sim0.15$ mags in $M_{K}$.}
\label{fig:HRdiagram}
\end{centering}
\end{figure}

Of the 244 targets presented here, 223 have parallax measurements from \textit{Hipparcos} or the second data release from \textit{Gaia}, listed in Table \ref{table:candidates}. The parallax measurements indicate that 31 of these systems lie within 125 pc, making them prime candidates for follow-up observation. We list these candidates in Table \ref{table:nearby_candidates}.

\begin{deluxetable*}{llccccc}
\tablecaption{Disk Candidates Within 125 Pc \label{table:nearby_candidates}}
\tablehead{\colhead{Zooniverse} & \colhead{WISEA} & \colhead{Distance(pc)} & \colhead{$V$ Magnitude} & \colhead{$T_{\mathrm{eff}}$ (K)} & \colhead{$T_{\mathrm{disk}}$ (K)} & \colhead{$L_{\mathrm{ir}}/L_{\mathrm{star}}$}}
\startdata
AWI00062lo & J003140.76-014737.3 & $109 \pm 6 $ & $7.07  \pm 0.01 $ & $10311_{- 62}^{+ 65}$ & $202_{-10}^{+11}$ & $8.5 (_{-0.58} ^{+0.54})  \times 10^{-5}$ \\
AWI0005yiv & J005926.26+400918.2 & $113 \pm 4 $ & $7.532 \pm 0.01 $ & $ 9169 \pm 57       $ & $ 59_{-17}^{+29}$ & $7.4 (_{-6.4}^{+140})     \times 10^{-4}$ \\
AWI000425z & J024755.37+553648.4 & $ 99 \pm 3 $ & $6.918 \pm 0.01 $ & $ 9490_{- 67}^{+ 66}$ & $131_{-43}^{+38}$ & $4.2 (_{-0.82}^{+4.7})    \times 10^{-5}$ \\
AWI0000phh & J025614.05+040254.2 & $125 \pm 21$ & $7.706 \pm 0.014$ & $11304_{-120}^{+122}$ & $215_{-14}^{+16}$ & $6.4 (_{-0.62}^{+0.60})   \times 10^{-5}$ \\
AWI0005yk3 & J030651.95+303136.8 & $ 73 \pm 3 $ & $7.061 \pm 0.011$ & $ 7640_{- 44}^{+ 46}$ & $243 \pm 5      $ & $7.5 (\pm 0.20)           \times 10^{-4}$ \\
AWI0005ykd & J032448.99+283908.6 & $106 \pm 6 $ & $7.121 \pm 0.01 $ & $ 9537_{- 60}^{+ 71}$ & $118 \pm 38     $ & $5.5 (_{-1.4}^{+11})      \times 10^{-5}$ \\
AWI0005zy4 & J032504.59+105835.0 & $116 \pm 5 $ & $7.294 \pm 0.011$ & $ 9493_{- 71}^{+ 69}$ & $182_{-51}^{+37}$ & $3.4 (_{-0.70}^{+0.85})   \times 10^{-5}$ \\
AWI00062iw & J032555.87-355515.1 & $100 \pm 6 $ & $6.385 \pm 0.009$ & $10160_{- 81}^{+ 86}$ & $216_{-21}^{+22}$ & $4.1 (_{-0.39}^{+0.42})   \times 10^{-5}$ \\
AWI0005ym9 & J035157.43+255955.4 & $121 \pm 9 $ & $7.68  \pm 0.01 $ & $ 9135_{- 48}^{+ 49}$ & $158_{-63}^{+44}$ & $3.9 (_{-1.1}^{+2.9})     \times 10^{-5}$ \\
AWI0005ymc & J040040.65+202447.8 & $119 \pm 5 $ & $8.05  \pm 0.01 $ & $ 8410 \pm 46       $ & $131_{-6 }^{+5 }$ & $5.7 (_{-0.35}^{+0.42})   \times 10^{-4}$ \\
AWI0005zz5 & J040238.47-004803.7 & $123 \pm 8 $ & $6.93  \pm 0.1  $ & $11167_{- 80}^{+ 81}$ & $176_{-24}^{+20}$ & $3.4 (_{-0.43}^{+0.47})   \times 10^{-5}$ \\
AWI0005ymi & J041249.03+193219.2 & $115 \pm 5 $ & $7.783 \pm 0.014$ & $ 6100 \pm 100      $ & $175_{-15}^{+14}$ & $2.3 (_{-0.19}^{+0.21})   \times 10^{-4}$ \\
AWI0005wcl & J045519.57+163712.9 & $109 \pm 8 $ & $7.17  \pm 0.01 $ & $10127_{- 80}^{+ 81}$ & $214_{-18}^{+21}$ & $5.1 (\pm 0.58)           \times 10^{-5}$ \\
AWI0005d88 & J083100.44+185806.0 & $ 91 \pm 4 $ & $7.418 \pm 0.012$ & $ 8319_{- 51}^{+ 47}$ & $ 45_{-10}^{+20}$ & $1.0(_{-0.96}^{+20})      \times 10^{-2}$ \\
AWI0000y1k & J111714.49-594610.8 & $ 76 \pm 1 $ & $7.15  \pm 0.01 $ & $ 8048_{- 51}^{+ 52}$ & $195_{-16}^{+17}$ & $1.3 (_{-0.096}^{+0.099}) \times 10^{-4}$ \\
AWI0005da8 & J112256.98-203731.7 & $114 \pm 5 $ & $7.466 \pm 0.011$ & $ 8835_{- 59}^{+ 64}$ & $132_{-48}^{+42}$ & $5.6 (_{-1.3}^{+8.7})     \times 10^{-5}$ \\
AWI00056ck & J121456.32-475654.6 & $117 \pm 5 $ & $8.097 \pm 0.011$ & $ 7889_{- 48}^{+ 47}$ & $150_{-8 }^{+7 }$ & $3.3 (_{-0.18}^{+0.20})   \times 10^{-4}$ \\
AWI00056i5 & J132026.77-491325.4 & $116 \pm 4 $ & $7.948 \pm 0.012$ & $ 8213_{- 55}^{+ 57}$ & $172_{-14}^{+13}$ & $1.3 (_{-0.093}^{+0.095}) \times 10^{-4}$ \\
AWI0000uji & J151147.67+101259.8 & $117 \pm 6 $ & $6.875 \pm 0.012$ & $10812_{- 99}^{+100}$ & $278_{-29}^{+38}$ & $3.7 (_{-0.49}^{+0.50})   \times 10^{-5}$ \\
AWI0000v1z & J152954.11+234901.6 & $114 \pm 4 $ & $7.585 \pm 0.011$ & $ 8576_{- 51}^{+ 49}$ & $ 47_{-13}^{+35}$ & $3.1 (_{-3.0}^{+160})     \times 10^{-3}$ \\
AWI00057qr & J164548.44-263858.1 & $108 \pm 6 $ & $7.02  \pm 0.011$ & $10863_{- 97}^{+104}$ & $193_{-10}^{+11}$ & $7.8 (\pm 0.54)           \times 10^{-5}$ \\
AWI00002yt & J172007.53+354103.6 & $101 \pm 2 $ & $8.272 \pm 0.012$ & $ 6800 \pm 100      $ & $196 \pm 30     $ & $9.0 (\pm 1.5)            \times 10^{-5}$ \\
AWI0005igi & J172452.23-185133.5 & $ 98 \pm 3 $ & $8.5   \pm 0.015$ & $ 6200 \pm 100      $ & $ 91_{-17}^{+13}$ & $2.3 (_{-0.71}^{+3.0})    \times 10^{-3}$ \\
AWI0005d5p & J180230.72+583738.4 & $ 89 \pm 4 $ & $6.858 \pm 0.01 $ & $ 9659 \pm 75       $ & $152_{-45}^{+35}$ & $3.3 (_{-0.44}^{+1.3})    \times 10^{-5}$ \\
AWI000621a & J205241.67-531624.8 & $ 63 \pm 1 $ & $7.663 \pm 0.011$ & $ 6300 \pm 100      $ & $ 47_{-10}^{+23}$ & $1.3 (_{-1.2}^{+21})      \times 10^{-2}$ \\
AWI0006222 & J210916.04-001405.6 & $103 \pm 9 $ & $6.659 \pm 0.01 $ & $ 9118_{- 55}^{+ 60}$ & $171_{-32}^{+24}$ & $5.4 (_{-0.51}^{+0.65})   \times 10^{-5}$ \\
AWI00019i2 & J221055.01+575629.4 & $112 \pm 4 $ & $7.42  \pm 0.01 $ & $ 8824_{- 54}^{+ 52}$ & $261_{-20}^{+21}$ & $8.4 (_{-0.79}^{+0.82})   \times 10^{-5}$ \\
AWI0000gjb & J224206.62-032824.4 & $112 \pm 12$ & $7.159 \pm 0.011$ & $11159_{- 98}^{+100}$ & $180_{-21}^{+19}$ & $3.8 (_{-0.40}^{+0.42})   \times 10^{-5}$ \\
AWI00062gs & J230533.05+145732.5 & $124 \pm 12$ & $6.762 \pm 0.01 $ & $11596 \pm 102      $ & $162_{-24}^{+19}$ & $3.3 (_{-0.40}^{+0.48})   \times 10^{-5}$ \\
AWI00062l3 & J235537.71+081323.7 & $113 \pm 8 $ & $6.818 \pm 0.01 $ & $11396_{-105}^{+109}$ & $269_{-29}^{+37}$ & $3.2 (_{-0.40}^{+0.41})   \times 10^{-5}$ \\
AWI00062h1 & J235746.21+112827.6 & $102 \pm 6 $ & $6.644 \pm 0.01 $ & $11092_{- 96}^{+ 92}$ & $205 \pm 27     $ & $2.6 (_{-0.35}^{+0.36})   \times 10^{-5}$ \\
\enddata
\end{deluxetable*}


We can also use these parallaxes and 2MASS photometry to construct an HR diagram for our disk candidates (shown in Figure \ref{fig:HRdiagram}). On this figure, point color corresponds to disk temperature, while point size corresponds to $L_{\mathrm{ir}}/L_{\star}$. A gray curve shows the zero-age main sequence \citep{2013ApJS..208....9P}. The point in gray in the lower left indicates typical uncertainties, $\sim 0.04$ mags in $\mathrm{(J-K_{s})}$ color and $\sim 0.15$ mags in $M_{\mathrm{K_{s}}}$. Although some stars have $K$-band excesses that shift their points to the right in this diagram, Figure \ref{fig:HRdiagram} reveals that none of the host stars with parallaxes has a main-sequence type later than roughly K7. Even with our ``drain-the-lake'' approach to searching for debris disks, the M dwarf disks remain hidden, at least in this sample.

\begin{figure}[htb]
\begin{centering}
\plotone{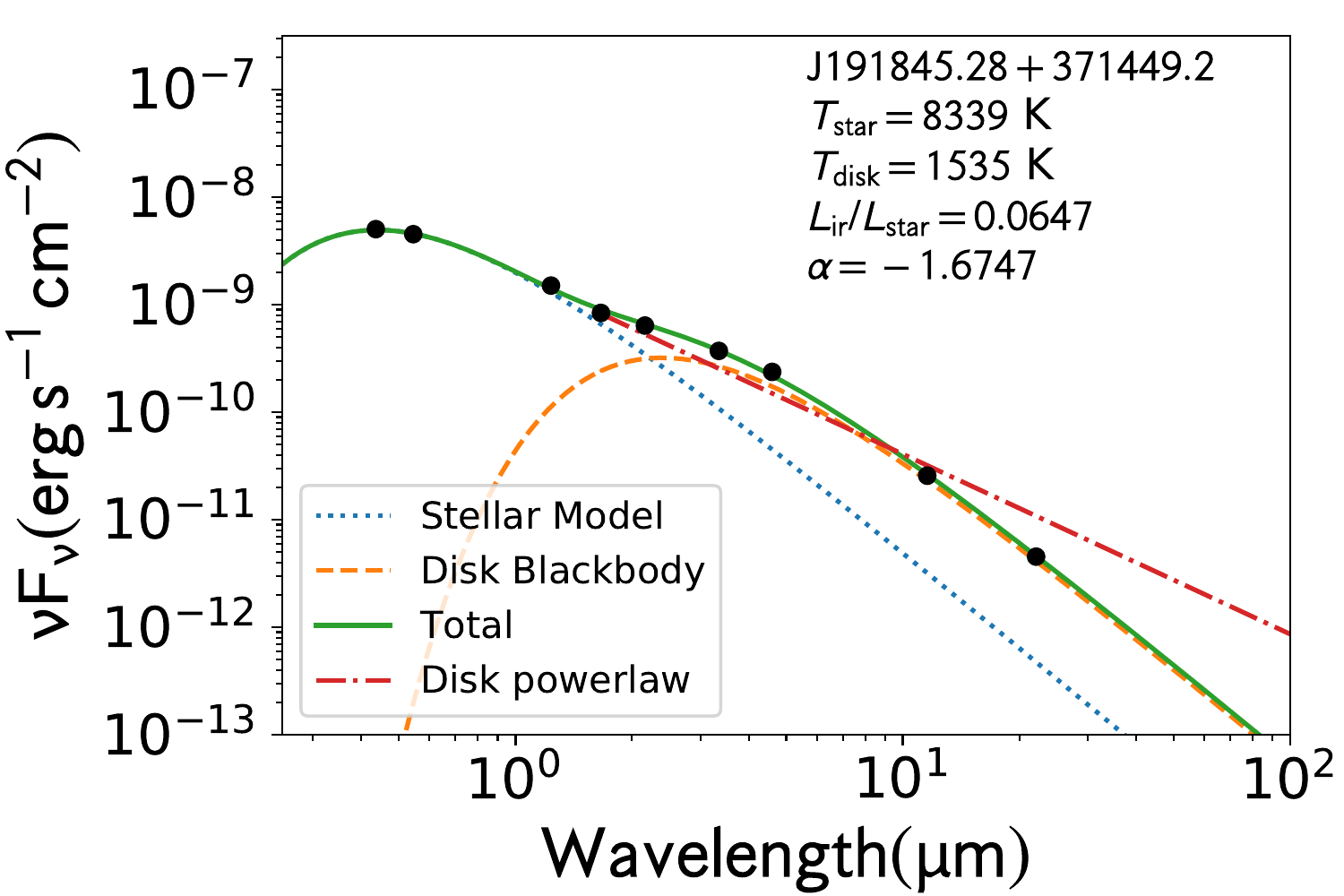}
\caption{SED for WISEA J191845.28+371449.2. This system is well fit by a stellar temperature blackbody with a $\sim 1535$ K disk blackbody with fractional infrared luminosity $\sim 0.065$, suggesting an extreme debris disk.}
\label{fig:extreme_SED}
\end{centering}
\end{figure}

The diagram also reveals some potentially interesting outliers. The yellow point near the center of the figure is WISEA J191845.28+371449.2. This star, listed in SIMBAD as an A2 at a distance of 573 pc, has excess flux in all three 2MASS bands and all four WISE bands, which is well modeled by a single-temperature blackbody at 1535 K with $L_{\mathrm{ir}}/L_{\star} = 0.0647$ (Figure \ref{fig:extreme_SED}).   The star lies roughly 0.7 mag to the right of the zero-age main sequence primarily because of its $K_s$ excess of 0.623 mag.  If this star is indeed a main sequence star with a debris disk as its SED fit suggests, this system would be an example of an ``extreme debris disk'' \citep{2015ApJ...805...77M, 2017AJ....153..165T}, i.e., a signpost of a recent giant impact. Given the lack of an additional mid-IR-driven excess, this object could also be an example of a precursor to a two component system with a hot debris disk \citep{2009ApJ...691.1896A}, where a gap has not yet been cleared between the two components. However, Gaia DR2 indicates that this object could be super-luminous for a spectral type of $\sim$A2 (the expected spectral type for the SED-fit stellar temperature), and optical spectroscopy indicates that this object could instead be a weak-lined Herbig Ae star (Bans et al. 2018, in prep). Additional follow-up with radio/sub-millimeter observations would be necessary to confirm whether this object hosts a warm debris disk, or instead has the expected substantial cold material reservoir of a Herbig Ae system.

\citet{2015ApJ...798...86Z} noted the high frequency of warm dust disks occurring around stars that were members of binary systems. 
\citet{2017AJ....153..257O} specifically searched for new comoving pairs and systems in the Tycho-Gaia Astrometric Solution \citep[TGAS;][]{2016A&A...595A...2G,2016A&A...595A...4L}. Out of 619,618 stars searched, they identified 8,472 stars as members of comoving pairs, and a further 2,134 as members of larger comoving systems. We note that 27 of our disk candidates presented here appear as members of comoving systems in \citet{2017AJ....153..257O} (listed in Table \ref{table:candidates}), out of 105 with parallax data sufficient for inclusion in the \citet{2017AJ....153..257O} survey. Twelve of these 105, or $11\% \pm 3\%$, are members of comoving pairs only, significantly higher than the overall rate of $1.37\%$ found by \citet{2017AJ....153..257O}. This significantly higher rate contributes further support to the hypothesis of \citet{2015ApJ...798...86Z}. 


We tested the likelihood of moving group membership for each of the new disk candidates we present here using BANYAN $\Sigma$ \citep{2018ApJ...856...23G}, and compared our targets with previous moving group membership determinations. We present the first 22 $\mu$m excess detection around J111714.49-594610.8, a known member of the Lower Centaurus Crux (LCC). Testing its kinematics with BANYAN $\Sigma$, however, indicates a 44.9\% probability of membership in LCC, and a 40.5\% probability of membership in Carina. We also note the first detection of an infrared excess around J140353.79-534628.3, a known Sco-Cen member \citep{2000MNRAS.313...43H} for which BANYAN $\Sigma$ yields an 88.9\% chance of membership in the Upper Centaurus-Lepus complex (UCL). We find that J164540.79-310226.6, a star with infrared excess previously detected by \citet{2014MNRAS.437..391C}, has a 96.9\% probability of membership in UCL. These determinations give us age benchmarks for these systems, allowing us to compare them to disk evolutionary models. Identification of disks in moving groups has yielded unexpected and valuable results with regard to theories of disk evolution, such as the identification of a primordial disk around an M dwarf in the Carina association, at an age 9 times greater than the $e$-folding time for primordial disk dissipation around solar-type stars \citep{2016ApJ...830L..28S, 2018MNRAS.476.3290M}.

Most of our published targets do not yet have spectroscopically-determined spectral types or luminosity classes, and many do not have measured distances. An ongoing spectroscopic follow-up campaign of DDOIs (Bans et al., in prep) will present a more complete and more detailed analysis of the distribution of spectral types of our objects. 

Between this paper, Paper 1, and \citet{2016ApJ...830L..28S}, Disk Detective has now published 215 previously-unidentified disk candidates. Of the 144 with either published spectral types or known parallaxes, the majority (110) are early-type main sequence stars. 125 of the 215 have disk temperatures $T_{\mathrm{disk}} < 300$ K, and disk temperatures range up to 1800 K. A majority of these disks (114) have fractional infrared luminosities $L_{\mathrm{ir}}/L_{\mathrm{star}} > 10^{-3}$, suggesting that these are likely primordial disks, per the criterion suggested by \citet{2011ARA&A..49...67W}; many of these do not appear in our previous HR diagram because they lack parallax measurements. The characteristics of this sample overall suggest that while designed to identify debris disks, Disk Detective also effectively locates new primordial disks. 




\section{Summary}
\label{sec:Conclusion}

In this paper, we presented the results of follow-up imaging of 261 Disk Detective Objects of Interest, determining whether background contaminants appeared and whether they significantly impacted the infrared excess around these objects observed with WISE. We find that background objects, while apparent in the images, significantly affect the observed excess at a rate of $\sim 6\%$. Combining these data with false-positive rates from classifications and literature review, we find that AllWISE should yield $\sim 21,600$ high quality disk candidates based on excess at 22 $\mu$m. Applying this result to other surveys, we estimate that $4-8\%$ of published disk candidates from high-quality surveys may have background objects in follow-up high-resolution imaging bright enough to significantly affect the detected excess. Based on our expected yield of disks from AllWISE, we found that the searches of \citet{2012MNRAS.427..343M}, \citet{2017MNRAS.471..770M}, and \citet{2016MNRAS.458.3479M} have lower-limit false positive rates greater than $70\%$. We considered the 175 disk candidates of \citet{2014ApJ...794..146T}, and found that the vast majority of these candidates would not be detected by Disk Detective due to insufficient signal-to-noise at W4. All thirteen targets in the \citet{2014ApJ...794..146T} search with W4 SNR sufficient for the Disk Detective methodology to apply were false-positive identifications after visual inspection.

We presented a sample of 244 disk candidates, vetted through visual inspection by citizen scientists and high-resolution follow-up imaging to refine the observed excess. Disk Detective has now published 215 newly-identified disk systems, of which 114 have fractional infrared luminosities indicative of primordial disks (either full protoplanetary or transitional). We find twelve of our disk candidates to be in comoving pairs, providing further support for the hypothesis of \citet{2015ApJ...798...86Z} that there is a causal relationship between a distant companion and a warm dusty debris disk. We identified one system, WISEA J191845.28+371449.2, as a likely ``extreme'' debris disk, based on its high fractional infrared luminosity. Thirty-one of these systems lie within 125 pc, including 27 debris disks. These nearby disk systems are good targets for adaptive-optics and coronagraphic imaging to directly image exoplanets in orbit around the host star.


High-resolution follow-up imaging can eliminate many false positives, but it will not eliminate not-yet-identified spectroscopic false-positive detections of primordial and debris disks (e.g. previously-unidentified dust-producing M giants, AGN, classical Be stars). A spectroscopic follow-up program to identify these sorts of false positives is ongoing. The results presented here should be used in conjunction with the results of that program (Bans et al., in prep) to determine the expected yield of AllWISE.



\acknowledgements

We thank the anonymous reviewer for providing comments that helped to improve the content and clarity of this paper, We acknowledge support from grant 14-ADAP14-0161 from the NASA Astrophysics Data Analysis Program and grant 16-XRP16\_2-0127 from the NASA Exoplanets Research Program. M.J.K. acknowledges funding from the NASA Astrobiology Program via the Goddard Center for Astrobiology. 

This publication makes use of data products from the Wide-Field Infrared Survey Explorer, which is a joint project of the University of California, Los Angeles, and the Jet Propulsion Laboratory (JPL)/California Institute of Technology (Caltech), and NEOWISE, which is a project of JPL/Caltech. WISE and NEOWISE are funded by NASA.

2MASS is a joint project of the University of Massachusetts and the Infrared Processing and Analysis Center (IPAC) at Caltech, funded by NASA and the NSF. 

The Digitized Sky Survey was produced at the Space Telescope Science Institute under U.S. Government grant NAG W-2166. The images of these surveys are based on photographic data obtained using the Oschin Schmidt Telescope on Palomar Mountain and the UK Schmidt Telescope. The plates were processed into the present compressed digital form with the permission of these institutions.

Funding for the SDSS and SDSS-II has been provided by the Alfred P. Sloan Foundation, the Participating Institutions, the National Science Foundation, the U.S. Department of Energy, the National Aeronautics and Space Administration, the Japanese Monbukagakusho, the Max Planck Society, and the Higher Education Funding Council for England. The SDSS Web Site is http://www.sdss.org/.

The SDSS is managed by the Astrophysical Research Consortium for the Participating Institutions. The Participating Institutions are the American Museum of Natural History, Astrophysical Institute Potsdam, University of Basel, University of Cambridge, Case Western Reserve University, University of Chicago, Drexel University, Fermilab, the Institute for Advanced Study, the Japan Participation Group, Johns Hopkins University, the Joint Institute for Nuclear Astrophysics, the Kavli Institute for Particle Astrophysics and Cosmology, the Korean Scientist Group, the Chinese Academy of Sciences (LAMOST), Los Alamos National Laboratory, the Max-Planck-Institute for Astronomy (MPIA), the Max-Planck-Institute for Astrophysics (MPA), New Mexico State University, Ohio State University, University of Pittsburgh, University of Portsmouth, Princeton University, the United States Naval Observatory, and the University of Washington.

This research has made use of the SIMBAD database, operated at CDS, Strasbourg, France. Some of the data presented in this paper were obtained from the Mikulski Archive for Space Telescopes (MAST). STScI is operated by the Association of Universities for Research in Astronomy, Inc., under NASA contract NAS5-26555. Support for MAST for non-HST data is provided by the NASA Office of Space Science via grant NNX13AC07G and by other grants and offices. This research has made use of the VizieR catalogue access tool, CDS, Strasbourg, France.

This work has made use of data from the European Space Agency (ESA) mission {\it Gaia} (\url{https://www.cosmos.esa.int/gaia}), processed by the {\it Gaia} Data Processing and Analysis Consortium (DPAC, \url{https://www.cosmos.esa.int/web/gaia/dpac/consortium}). Funding for the DPAC has been provided by national institutions, in particular the institutions participating in the {\it Gaia} Multilateral Agreement.

The Robo-AO system was developed by collaborating partner institutions, the California Institute of Technology and the Inter-University Centre for Astronomy and Astrophysics, and with the support of the National Science Foundation under Grant Nos. AST-0906060, AST-0960343 and AST-1207891, the Mt. Cuba Astronomical Foundation and by a gift from Samuel Oschin. C.B. acknowledges support from the Alfred P. Sloan Foundation.

IRAF is distributed by the National Optical Astronomy Observatory, which is operated by the Association of Universities for Research in Astronomy (AURA) under a cooperative agreement with the National Science Foundation. PyRAF is a product of the Space Telescope Science Institute, which is operated by AURA for NASA. This research made use of ds9, a tool for data visualization supported by the Chandra X-ray Science Center (CXC) and the High Energy Astrophysics Science Archive Center (HEASARC) with support from the JWST Mission office at the Space Telescope Science Institute for 3D visualization. Resources supporting this work were provided by the NASA High-End Computing (HEC) Program through the NASA Center for Climate Simulation (NCCS) at Goddard Space Flight Center.

\facilities{FLWO:2MASS, CTIO:2MASS, WISE, PO:1.5m (Robo-AO), Du Pont (RetroCam)}

\software{IRAF \citep{1993ASPC...52..173T}, PyRAF, AstroPy \citep{2013A&A...558A..33A}, NumPy \citep{van2011numpy}, SciPy \citep{jones_scipy_2001}, Matplotlib \citep{Hunter:2007}, pandas \citep{mckinney}, emcee \citep{2013ascl.soft03002F}}

\appendix

\section{Changes to the Website Classification Scheme after Paper 1}

Since Paper 1, we have made two key changes to the website classification setup; we changed the scheme for retiring subjects, and we corrected the online photometry to account for saturation effects at W1 and W2. 

\subsection{New Retirement Scheme}
\label{sec:retirement}
Visitors to the diskdetective.org site (``users'') view ``flipbooks'' showing several images of the same source at different wavelengths.  After they view the flipbooks, users answer a question, ``What best describes the object you see?'', by clicking on one or more of six buttons, labeled ``Multiple objects in the Red Circle'', ``Object Moves off the Crosshairs'', ``Extended beyond circle in WISE Images'' ``Empty Circle in WISE images'', ``Not Round in DSS2 or 2MASS images'' and ``None of the Above/Good Candidate''.  With the exception of the ``None of the Above'' option, the user can choose more than one description per flipbook.  After at least one of these classification buttons is chosen, a button labeled ``Finish'' becomes active; clicking this button records the user's choices and causes the next flipbook to appear.

To utilize this classification data requires a ``retirement scheme,'' a set of rules for deciding when a given subject has received enough classifications that we can be reasonably sure of whether or not it is a good candidates.  Prior to July 23, 2015, we used a very simple retirement scheme; we retired all subjects after 15 classifications.  However, we realized that certain kinds of sources did not require so many independent classifications to make a confident decision about their nature. So on July 23, 2015, we put in place a new retirement scheme that allows us to progress more rapidly through the data.

To develop this new scheme, we experimented with several possible retirement rules by applying them to a set of subjects that already had fifteen classifications, to see if they would alter the final classification.  We found that users were especially reliable at classifications as ``Multiple objects in the Red Circle'', or ``Not Round in DSS2 or 2MASS images''.  Even if we used only the first five classifications of subjects in these categories, it did not change their ultimate classification.  So we chose to implement a retirement rule that retires the subjects as either ``Multiple objects..." or ``Not Round..." when four out of the first five classifications are either ``multi" or ``oval" respectively.  From the first year of classifications we know that almost 45\% of our sources fit in one of these two categories.  So we expected a substantial increase in efficiency from this rule, and indeed, though the number of active participants in the project remained stable roughly since July 2014, we have seen a noticeable increase in the retirement rate since implementing the new rule, so that 26\% of our 278,121 subjects have now been retired. 

\subsection{Removal of WISE 1 Dropouts}
\label{sec:dropouts}

A second improvement we made to the website was that we removed a list of problematic subjects from the online classification process.  As \citet{2014ApJS..212...10P} and others have noted, bright sources can saturate the WISE detectors, causing systematic errors in the WISE photometry.  These errors are worse at the WISE W1 and W2 bands.  Since we chose objects for our input catalog using [W1]-[W4] colors, these photometric errors caused us to include some objects in our input catalog that had no true excess at W4 (22$\mu$m), only false deficits at W1.  Using the saturation corrections in \citet{2014ApJS..212...10P}, we found a list of 279 subjects that had been included in our search incorrectly because of saturation errors in their W1 photometry. On August 3, 2015, we removed these sources from the vetting process.

\section{Deriving Brightness Thresholds for Contaminants}
\label{sec:appendix}

In this section, we derive a minimum brightness for background objects to produce a false positive excess detection in our system. We also show how these brightnesses propagate to the $i$, $Y$ and $H$ bands we use in our follow-up observations.

\subsection{Minimum Contaminant Brightness in W4}

Let us suppose that we have a target star with \textit{no} circumstellar dust, whose spectrum is accurately approximated by the Rayleigh-Jeans law in the WISE bands. We will denote this target star's intrinsic magnitudes as $m_{t}$, and its intrinsic flux as $f_{m,t}$. Let us also suppose a background contaminant whose intrinsic magnitudes will be denoted $m_{c}$ and whose flux will be denoted $f_{m,c}$. This background contaminant lies substantially inside the W4 PSF half-width at half-maximum, such that the observed W4 magnitude $[W4]_{obs}$ is the magnitude of the combined light from the target and contaminant:
\begin{equation}
    f_{W4,obs} = f_{W4,t} + f_{W4,c},
\end{equation}
\noindent
but outside the W1 PSF HWHM, such that $[W1]_{obs}$ = $[W1]_{t}$.

Assuming that our target star has $[W1_{t}] - [W4_{t}] = 0$, we wish to know the minimum brightness $W4_{c}$ such that $[W1_{obs}] - [W4_{obs}] \geq 0.25$.

Making substitutions suggested by the above equalities, we have

\[
    W4_{t} - W4_{obs} \geq 0.25,
\]

Using the equation for flux/magnitude conversion, this becomes

\begin{equation}
    f_{W4,c} \geq 0.258925 f_{W4,t}.
\end{equation}

Converting this back into magnitude differences, we have

\begin{equation}
    W4_{c} - W4_{t} \leq 1.467
\end{equation}

We thus show that contaminants more than $\sim 1.5$ magnitudes dimmer than the target star in W4 will not produce a significant enough excess for the star to become a false positive entry in the Disk Detective Input Catalog. As such, with the greater depth probed by high-resolution imaging, we must quantitatively assess whether a detected background contaminant will produce a false positive, rather than qualitatively assessing it, as was done with the Web site-based classifications. \citep{2016IAUS..314..159B}

\subsection{Applying the Minimum Brightness to the i, Y, and H Bands}

To determine if a potential background object is bright enough to produce a false positive excess in W4, we must know the difference in magnitude between such a contaminant and the target in the bands in which we conducted high-resolution follow-up observations. We assume for this exercise that the target's optical and near-IR SED is approximately identical to an idealized Vega---i.e. it is of zeroth magnitude in all bands. This sets the contaminating magnitude limit of 1.467. Below, we give examples of three typical contaminants, each with a W4 magnitude of 1.467: a background M dwarf and two different power-law spectra. Table \ref{table:colors} lists the $i$, $Y$, and $H$ magnitudes for these objects. 

\begin{deluxetable}{cccc}
\tablecaption{W4 Colors of Selected Background Object Types \label{table:colors}}
\tablehead{\colhead{} & \multicolumn{3}{c}{Colors} \\ \colhead{Characteristics} & \colhead{$i - W4$} & \colhead{$Y_{C} - W4$} & \colhead{$H_{C} - W4$}}
\startdata
M dwarf: $T_{\mathrm{eff}} = 3300 \mathrm{K}, \log(g) = 5.0$ & 3.545 & 1.896 & 0.801 \\
Class I YSO: $\lambda f_{\lambda} \sim \lambda^{0}$ & 2.94 & 2.05 & 1.82 \\
ULIRG: $\lambda f_{\lambda} \sim \lambda^{2}$ & 10.25 & 8.67 & 7.48 \\
\enddata
\end{deluxetable}

\subsubsection{False Positive due to a Background M dwarf}
\label{sec:cont_mdwarf}

The initial mass function of the neighborhood peaks at a spectral type of M2-M3.5, which corresponds to a stellar effective temperature of $T_{\mathrm{eff}} \simeq 3250-3400$ K. Accordingly, we select a model M dwarf atmosphere with $T_{\mathrm{eff}} = 3300$ K and $\log(g) = 5.0$ from the BT-Settl package of model atmospheres \citep{Baraffe2015} as our contaminating M dwarf. The BT-Settl models have pre-computed magnitudes for many filter systems, which we use here to compute colors. This model has a [H]-[W4] value of 0.801, corresponding to an $H$ band contaminant delta-magnitude of 2.268 magnitudes. The [$i$]-[W4] value for this model is 3.545 magnitudes, yielding a magnitude difference in Robo-AO data of 5.012 magnitudes.

\subsubsection{False Positive due to Background Sources with Power Law Spectra}


Some objects are reasonably represented with a power-law spectrum $\lambda f_{\lambda} \sim \lambda^{\alpha}$. For these objects, colors can to first order be approximated as 

\[
[m_{\lambda_{1}}] - [m_{\lambda_{2}}] \sim 2.5\log_{10}\left(\frac{f_{0,\lambda_{1}}}{f_{0,\lambda_{2}}}\right) - 2.5(\alpha-1)\log_{10}\frac{\lambda_{1}}{\lambda_{2}}.
\]

\noindent
For the colors of interest in this paper, this corresponds to:

\[
[\mathrm{H}] - [\mathrm{W4}] \sim 1.817 + 2.831\alpha,
\]

\[
[\mathrm{Y}] - [\mathrm{W4}] \sim 2.048 + 3.313\alpha,
\]

\noindent
and

\[
[i] - [\mathrm{W4}] \sim  2.940 + 3.654\alpha.
\]

Young stellar objects are defined by the value of $\alpha$---a Class I YSO has slope $\alpha = 0$ \citep{1987IAUS..115....1L, 1995ApJS..101..117K}; this spectrum also roughly approximates a heavily-reddened early-type star. If such an object were in the background of our images and bright enough to produce a false-positive, the delta-magnitude in the H band would be 3.284 and the Y band delta-magnitude would be 3.515. These could both likely be detected in our images. The $i$-band delta-magnitude would be 4.407, also likely detectable.

Also of interest is the case where $\alpha = 2$. This roughly corresponds to the SED of a luminous or ultra-luminous infrared galaxy, or (U)LIRG, which has an AGN component \citep{2008A&A...484..631V}. In this case, however, the H-band delta-magnitude would be 8.946 and the Y-band delta-magnitude would be 10.141, while the $i$-band delta-magnitude would be 11.715. These are all clearly undetectable in our follow-up image data, and are thus treated instead with the prescription of \citet{Papovich2004} in Section \ref{sec:ulirgs}.

\section{Comments on Selected Disk Candidates}
\label{sec:appendix_candidates}

Below are brief comments on selected disk candidates presented in Section \ref{sec:new_candidates}. Having noted previous identifications as disk candidates in Section \ref{sec:new_candidates}, we primarily discuss characteristics of the subject's appearance, either in the images used on the Disk Detective website, or in the follow-up images.

\begin{itemize}

\item \textbf{J021327.01+421923.3} This system, which was previously identified by \citet{2016ApJS..225...15C,2017MNRAS.471..770M}, exhibits slight extension in the W4 image. 

\item \textbf{J023720.84+395345.8} This system is a known spectroscopic binary \citep{1981A&AS...44...59H}. The companion, an early-G main sequence star based on the binary mass function, has a projected angular separation of $<<1''$, making it undetectable as a separate component in follow-up data. The SED indicates no significant effect on the 22 micron excess.

\item \textbf{J025926.83+593531.6} We recover this system, a spectroscopic binary \citep{2009ApJS..180..117A} which was previously identified as a source in the W5 region by \citet{2008ApJ...688.1142K}. The secondary component has a minimum mass of 1.7326 $R_{\odot}$ based on the binary mass function, suggesting a possible A7V star. The projected angular separation is $<<1''$, making it undetectable as a separate component in follow-up data. The SED indicates no significant effect on the 22 micron excess. 

\item \textbf{J030854.20-185809.1} This system, a known A0V system, is best fit as a two-stellar-component system with a K dwarf component, as well as a 349 K disk. Further observation is necessary to confirm this additional component.

\item \textbf{J034400.28+243324.6} This $\delta$ Scuti variable, which appears in \citet{2016ApJS..225...15C,2017MNRAS.471..770M}, exhibits slight extension at W4.

\item \textbf{J041517.47+505124.0} This system, which appears in \citet{2016MNRAS.458.3479M}, has been identified as a Be star previously \citep{1943ApJ....98..153M} and more recently as having H-alpha in emission \citep{1999A&AS..134..255K}. Further spectroscopic follow-up is necessary to determine the nature of this emission (i.e. whether the target is a classical Be star, rather than a debris disk host).

\item \textbf{J043521.12-081730.0} This object exhibits slight extension at W4.

\item \textbf{J051143.75+122012.5} This system exhibits slight extension at W4. 

\item \textbf{J051328.63-043910.6} This object is a component of a binary system. Its companion is $35''$ away from the source, too distant to affect the observed excess.

\item \textbf{J052331.01-010423.6} This star exhibits eclipses from circumstellar material, per \citet{2017MNRAS.471..740O}.

\item \textbf{J053707.15+603636.4} This object, which appears in \citet{2016ApJS..225...15C} exhibits slight extension at W4.

\item \textbf{J054330.38+251724.4} This system has previously been identified as an H-alpha emitter \citep{1999A&AS..134..255K}. Further spectroscopic follow-up is necessary to determine the nature of this emission.

\item \textbf{J054733.26+521144.5} This system has previously been identified as an H-alpha emitter \citep{1950ApJ...112...72M}. Further spectroscopic follow-up is necessary to determine the nature of this emission.

\item \textbf{J111714.49-594610.8} This target is a member of the Lower Centaurus Crux (LCC) of the Sco-Cen OB2 association, per \citet{2000MNRAS.313...43H}. Testing its kinematics with BANYAN $\Sigma$ yields a 44.9\% membership probability in LCC, and a 40.5\% membership probability in Carina.

\item \textbf{J111925.92-301922.9 and J165204.85+145827.2} We report the first detection of a disk around the $\alpha^{2}$ CVn variable AWI0005cwg. We also note that AWI00004o8, identified in Paper 1, is also an $\alpha^{2}$ CVn variable.

\item \textbf{J114336.83-802900.5} This target exhibits slight extension at W4.

\item \textbf{J132026.77-491325.4} This target is a known disk-hosting member of Sco-Cen \citep{2012ApJ...756..133C}. Re-evaluation with BANYAN $\Sigma$ yields an 88.7\% probability of membership in LCC.

\item \textbf{J134909.18-541342.3} This target is a known disk-hosting member of Sco-Cen \citep{2012ApJ...756..133C}. Re-evaluation with BANYAN $\Sigma$ yields a 42.1\% membership probability in LCC, and a 49.5\% membership probability in Upper Centaurus-Lepus (UCL).

\item \textbf{J144458.63-280251.9} This target is a known multiple system \citep{2011AJ....141...45H} with an excess first detected by \citet{2014MNRAS.437..391C}. We do not resolve both components of the binary system.

\item \textbf{J173254.69+404312.3} This target exhibits very slight extension at W4.

\item \textbf{J173832.90+425112.9} This target is a known Cepheid variable.

\item \textbf{J183311.41+025439.0} This object shows a faint background object in DSS2 survey data that does not appear in Robo-AO observations.

\item \textbf{J185211.39+102422.6} This object shows a faint background object in DSS2 survey data that does not appear in Robo-AO observations.

\item \textbf{J190901.24+110641.3} This target exhibits H$\alpha$ in emission \citep{1999A&AS..134..255K}.

\item \textbf{J192136.46+220744.7} This object shows a faint background object in DSS2 survey data that does not appear in Robo-AO observations.

\item \textbf{J192437.52+563454.9} This object shows a very faint background object in DSS2 survey data, which does not appear in the Robo-AO data for the system.

\item \textbf{J210144.07+521717.6} This object is an emission-line star \citep{1999A&AS..134..255K}.

\item \textbf{J212952.96+525601.9} This object exhibits a slight asymmetrical extension at W4.

\item \textbf{J215947.70-593411.9} This target is a known $\delta$ Scuti variable \citep{2000A&AS..144..469R}.

\item \textbf{J230112.67-585821.9} We note slight extension at W4.

\end{itemize}

\end{document}